\documentclass{article}

\usepackage{fullpage}
\usepackage{textcomp}

\usepackage{subcaption}
\usepackage{enumerate}
\usepackage{amsmath}
\usepackage{amsfonts}
\usepackage{graphicx}
\usepackage{hyperref}

\newtheorem{lemma}{Lemma}
\newtheorem{theorem}[lemma]{Theorem}

\newtheorem{definition}[lemma]{Definition}

\title{On the Separability of Stochastic Geometric Objects, with Applications}
\author{
  Jie Xue\footnote{Dept. of Computer Science and Engg., Univ. of Minnesota --- Twin Cities, 4-192 Keller Hall, 200 Union St. SE, Minneapolis, MN 55455, USA}\\
  \texttt{xuexx193@umn.edu}
  \and
  Yuan Li\footnotemark[1]\\
  \texttt{lixx2100@umn.edu}
  \and
  Ravi Janardan\footnotemark[1]\\
  \texttt{janardan@umn.edu}
}
\date{}
\begin{document}

\maketitle

\begin{abstract}
In this paper, we study the linear separability problem for stochastic geometric objects under the well-known unipoint/multipoint uncertainty models. Let $S=S_R \cup S_B$ be a given set of stochastic bichromatic points, and define $n = \min\{|S_R|, |S_B|\}$ and $N = \max\{|S_R|, |S_B|\}$.
We show that the {\it separable-probability} (SP) of $S$ can be computed in $O(nN^{d-1})$ time for $d \geq 3$ and $O(\min\{nN \log N, N^2\})$ time for $d=2$, while the {\it expected separation-margin} (ESM) of $S$ can be computed in $O(nN^{d})$ time for $d \geq 2$.
In addition, we give an $\Omega(nN^{d-1})$ {\it witness-based lower bound} for computing SP, which implies the optimality of our algorithm among all those in this category.
Also, a hardness result for computing ESM is given to show the difficulty of further improving our algorithm.
As an extension, we generalize the same problems from points to general geometric objects, i.e., polytopes and/or balls, and extend our algorithms to solve the generalized SP and ESM problems in $O(nN^{d})$ and $O(nN^{d+1})$ time, respectively.
Finally, we present some applications of our algorithms to stochastic convex-hull related problems.
\end{abstract}

\section{Introduction}\label{sec:introduction}
Linear separability describes the property that a set of $d$-dimensional bichromatic (red and blue) points can be separated by a hyperplane such that all the red points lie on one side of the hyperplane and all the blue points lie on the other side.
This problem has been well studied for years in computational geometry, and is widely used in machine learning and data mining for data classification.
However, existing linear-separation algorithms require that all the input points must have fixed locations, which is rarely true in reality due to imprecise sampling from GPS, robotics sensors, or some other probabilistic systems.
It is therefore essential to study the conventional linear separability problem under uncertainty.

In this paper, we study the linear separability problem  under two different uncertainty models, i.e., the 
{\it unipoint} and {\it multipoint} models \cite{agarwal2014convex}.
In the former, each stochastic data point has a fixed and known location, and has a positive probability to exist at that location;
whereas in the latter, each stochastic data point occurs in one of discretely-many possible locations with a known probability,
and the existence probabilities of each point sum up to at most 1 to allow for its absence.
Our focus is to compute the {\it separable-probability} (SP) and the {\it expected separation-margin} (ESM) for a given set of bichromatic stochastic points (or general geometric objects) in $\mathbb{R}^d$ for $d \ge 2$, where the former is the probability that the existent points (or objects) are linearly separable, and the latter is the expectation of the separation-margin of the existent points (or objects).
(See Section~\ref{subsec2.2.1} for a detailed and formal definition of the latter.)

The brute-force approach of enumerating all possible instances takes exponential runtime, and therefore we propose, in this paper, novel algorithms that carefully compute SP and ESM in a much more efficient way.
Furthermore, we show that our approach is highly extensible and can solve many other related problems defined on other types of objects or on multiple colors.
To summarize, our main contributions are:
\begin{enumerate}
	\item[(1)] We propose an $O(nN^{d-1})$-time algorithm, which uses linear space, for solving the SP problem when given a set of bichromatic stochastic points in $\mathbb{R}^d$, $d \ge 3$.
	(The runtime is $O(\min\{N^2, nN\log N\})$ for $d = 2$.)
	We also show an $\Omega(nN^{d - 1})$ lower bound for all witness-based algorithms, which implies the optimality of our algorithm among all witness-based methods for $d \ge 3$. (See Section~\ref{sec:separable-prob}.)
	
	\item[(2)] We show that the ESM of the above dataset can be computed in $O(nN^d)$ time for $d \geq 2$, using linear space.
	A hardness result is also given to show the total number of distinct possible separation-margins is $\Theta(nN^d)$, which implies that it may be difficult to achieve a better runtime. (See Section~\ref{sec:separation-margin}.)
	
	\item[(3)] We extend our algorithms to compute the SP and the ESM for datasets containing general stochastic geometric objects, such as polytopes and/or balls.
	Our generalized algorithms solve the former problem in $O(nN^d)$ time, and the latter in $O(nN^{d+1})$ time, using linear space.  (See Section~\ref{sec:extension:shape}.)
	
	\item[(4)] We provide some applications of our algorithms to stochastic convex-hull (SCH) related problems.
	Specifically, by taking advantage of our SP algorithm, we give a method to compute the SCH membership probability, which matches the best known bound but is more direct.
	Also, we consider some generalized versions of this problem and show how to apply our separability algorithms to solve them efficiently. (See Section~\ref{sec:application}.)
\end{enumerate}

To provide a smooth flow for the paper, all proofs and some details are deferred to Appendix~\ref{append.proof}.

\subsection{Related work}\label{sec:related_work}
The study of computational geometry problems under uncertainty is a relatively new topic, and has attracted a lot of attention; see \cite{aggarwal2009survey} and \cite{dalvi2009probabilistic} for two surveys. Different uncertainty models and related problems have been investigated in recent years.  See \cite{agarwal2013nearest, agarwal2009indexing, agarwal2012range, deBerg2014separability, evans2008guaranteed, loffler2009data, loffler2010largest, zhang2012nearest} for example. The unipoint/multipoint uncertainty model, which we use in this paper, was first defined in \cite{agarwal2014convex, suri2013most}, and has been applied in many recent papers. For instance, in \cite{kamousi2011stochastic}, Kamousi et al. studied the stochastic minimum spanning tree problem, and computed its expected length.
Suri et al. investigated the most likely convex hull problem over uncertain data in \cite{suri2013most}; the similar topic was revisited by Agarwal et al. in \cite{agarwal2014convex} to compute the probability that a query point is inside the uncertain hull. In \cite{suri2014most}, Suri and Verbeek studied the most likely Voronoi Diagram (LVD) in $\mathbb{R}^1$ under the unipoint model, and the expected complexity of LVD was further improved by Li et al. in \cite{li2015arrangement}, who explored the stochastic line arrangement problem in $\mathbb{R}^2$. In \cite{agrawal2015skyline}, Agrawal et al. proposed efficient algorithms for the most likely skyline problem in $\mathbb{R}^2$ and gave NP-hardness results in higher dimensions.

Recently, in \cite{deBerg2014separability}, de Berg et al. studied the separability problem given a set of bichromatic imprecise points in $\mathbb{R}^2$ in a setting that each point is drawn from an imprecision region.

Very recently, in an unpublished manuscript \cite{martin2015seperability},
one of our proposed problems, the SP computing problem, was independently studied
by Fink et al. under the same uncertainty model, and similar results were obtained, i.e., an $O((n+N)^d) = O(N^d)$-time and $O(n+N) = O(N)$-space algorithm for computing the SP of a given set of bichromatic stochastic points in $\mathbb{R}^d$.
In fact, before the final step of the improvement, the time bound achieved was $O(N^d \log N)$, and then {\it duality} \cite{deBerg_CG_book} and {\it topological sweep} \cite{Edelsbrunner:1986:topological_sweep} techniques were applied to eliminate the $\log$ factor.
On the other hand, our algorithm runs initially in $O(nN^{d-1} \log N)$ time using linear space, and can be further improved to $O(nN^{d-1})$ runtime by using the same techniques. (A careful discussion will be given in Section~\ref{sec:separable-prob}.)
Note that the algorithm in \cite{martin2015seperability} always runs in $\Theta(N^d)$ time (no matter how small $n$ is) and, more importantly, this runtime appears to be intrinsic: all possible $d$-tuples of (distinct) points have to be enumerated in order to correctly compute the SP.
Our time bound matches the bound in \cite{martin2015seperability} when $n = \Theta(N)$.
However, when $n \ll N$, our method is much faster, and, in fact, this is usually the case in many real classification problems in machine learning and data mining.

In terms of how to solve the problem, Fink et al.'s method is very different from ours. 
Their computation of SP relies on an additional dummy anchor point, and based on this point, the probability is computed in an inclusion-exclusion manner.
On the other hand, our method solves the problem more directly: it does not introduce any additional points and the SP is eventually computed using a simple addition principle.
Furthermore, our algorithm can be easily extended to solve many generalized problems (e.g., multiple colors, general geometric objects, etc.)

\subsection{Basic notations and preliminaries} \label{sec:basic-notations}
Throughout this paper, the basic notations we use are the following.
We use $S = S_R \cup S_B$ to denote the given stochastic bichromatic dataset, where
$S_R$ (resp. $S_B$) is a set of red (resp. blue) stochastic points (or general geometric objects in Section~\ref{sec:extension:shape}).
The notations $n$ and $N$ are used to denote the sizes of the smaller and larger classes of $S$ respectively, i.e., $n = \min\{|S_R|,|S_B|\}$ and $N = \max\{|S_R|,|S_B|\}$,
and $d$ is used to denote the dimension.
In this paper, we always assume that $d$ is a constant.
When we need to denote a normal bichromatic dataset (without considering the existence probabilities), we usually use the notation $T = T_R \cup T_B$.
The coordinates of a point $x \in \mathbb{R}^d$ are denoted as $x^{(1)}, x^{(2)}, \dots, x^{(d)}$.
If $T$ is a dataset in $\mathbb{R}^d$ and $U$ is some linear subspace of $\mathbb{R}^d$, we use $T^U$ to denote a new dataset in the space $U$, which is obtained by orthogonally projecting $T$ from $\mathbb{R}^d$ onto $U$.

We say a set of bichromatic points is {\it strongly separable} iff there exists a hyperplane, $h$, so that all the red points strictly lie on one side of $h$ while all the blue points strictly lie on the other side.
Also, we can define the concept of {\it weakly separable} similarly, except that we allow points to lie on the hyperplane.
A hyperplane that strongly (resp., weakly) separates a set of bichromatic points is called a {\it strong} (resp., {\it weak}) {\it separator}.
The following is a fundamental result we will use in various places of this paper.
\begin{theorem} \label{th1}
	Suppose $T = T_R \cup T_B$ is a set of bichromatic points in $\mathbb{R}^d$. $T$ is strongly separable by a hyperplane 
	iff 
	$\mathcal{CH}(T_R) \cap \mathcal{CH}(T_B) = \emptyset$,
	where $\mathcal{CH}(\cdot)$ denote the convex hull of the indicated point-set.
\end{theorem}

\section{Separable-probability}\label{sec:separable-prob}
The separable-probability (SP) of a stochastic bichromatic dataset $S = S_R \cup S_B$ in $\mathbb{R}^d$ refers to the probability that the existent points in $S$ (obtained by a random experiment) are (strongly) separable.
For simplicity, we describe here the details of our algorithm under the unipoint model only.
The generalization of our algorithm to the multipoint model is quite straightforward and we discuss it in Appendix~\ref{sec:multipoint}.

\begin{figure}[htpb]
    \centering
    \includegraphics{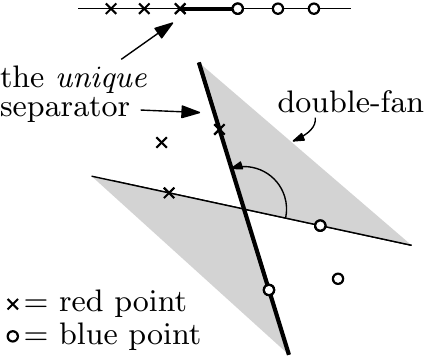}
    \caption{Illustrating the unique separator we choose for separable instances in $\mathbb{R}^1$ and $\mathbb{R}^2$.}
    \label{fig:intro_to_extreme}
\end{figure}

Trivially, given a dataset $S$, its SP can be computed by simply enumerating all the $2^{|S|}$ possible instances of $S$ and summing up the probabilities of the separable ones, which takes exponential time.
In order to solve the problem more efficiently than by brute-force, one has to categorize all the separable instances of $S$ into a reasonable number of groups such that the sum of the probabilities of the instances in each group can be easily computed.
A natural approach is to charge each separable instance to a {\it unique} separator, and use that as the key to do the grouping.
The uniqueness requirement here is to avoid over-counting.
In addition, all these separators should be easy to enumerate and the sum of the probabilities of those separable instances charged to each separator should be efficiently computable.
In $\mathbb{R}^1$ and $\mathbb{R}^2$, this is easy to achieve.
For example, in $\mathbb{R}^1$, given a separable instance, all the possible separators form a segment, and we can choose the leftmost endpoint as the unique separator; in $\mathbb{R}^2$, all the possible separators of a separable instance form a double-fan, and we can choose the most counterclockwise one, which goes through exactly one red and one blue point, as the unique separator. (See Figure~\ref{fig:intro_to_extreme} for an illustration.)
It is easy to see that, with the separators chosen above, the SP of the stochastic dataset can be easily computed by considering the sum of the probabilities of the instances charged to each such separator.
However, to define such a separator for a separable instance beyond $\mathbb{R}^2$ turns out to be hard and challenging. 
To solve this problem, we define an important concept called {\it extreme separator} in Section~\ref{subsec2.1.1}, and apply this concept to compute the SP of $S$ in Section~\ref{subsec2.1.2}.

\begin{figure}[htpb]
    \centering
    \includegraphics{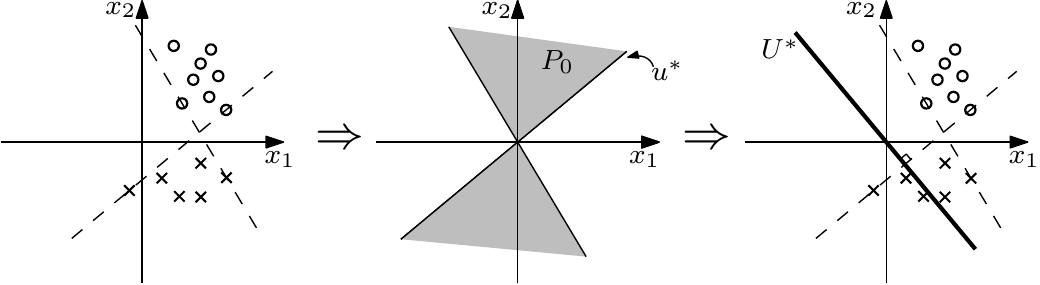}
    \caption{Illustrating $U^*$ in $\mathbb{R}^2$. Note that $P_1$ is not shown to avoid confusion.}
    \label{fig:p0p1}
\end{figure}
\begin{figure}[htpb]
    \centering
    \includegraphics{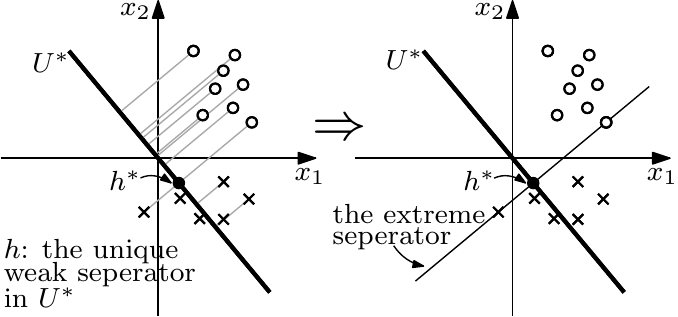}
    \caption{Illustrating the extreme separator in $\mathbb{R}^2$.}
    \label{fig:extreme}
\end{figure}


For convenience, we assume that the points given in $S$ have the \textit{strong general position} property (SGPP), which is defined as follows.
Let $I=\{i_1,i_2,\dots,i_{|I|}\}$ be any subset of the index set $\{1,2,\dots,d\}$ where $i_1<i_2<\dots<i_{|I|}$.
We define a projection function $\phi_I:\mathbb{R}^d \rightarrow \mathbb{R}^{|I|}$ as
$\phi_I(x) = (x^{(i_1)},x^{(i_2)},\dots,x^{(i_{|I|})})$.
Also, for any $X \subseteq \mathbb{R}^d$, we define
$\Phi_I(X)=\{\phi_I(x):x \in X\}$.

Let $T$ be a set of points in $\mathbb{R}^d$.
When $d \leq 2$, we say $T$ has SGPP iff it is in general (linear) position, i.e., affinely independent.
When $d \geq 3$, we say $T$ has SGPP iff \\
\textbf{1)} $T$ is in general (linear) position; \\
\textbf{2)} $\Phi_J(T)$ has SGPP for $J = \{3,4,\dots,d\}$.\\
{\bf Remark.} Note that this assumption is actually not stronger than the conventional general position assumption, since one can easily apply linear transformations to make a set of points in general position have SGPP (the separability of a dataset is invariant under non-singular linear transformations). In fact, a randomly-generated non-singular $d \times d$ matrix $\mathcal{A}$ can transfer a set of points in general position into another set having SGPP with a probability almost 1. Though this method works well in practice, we give, in Appendix~\ref{append:compute_matrix_A}, a deterministic algorithm that guarantees to output a valid $\mathcal{A}$ in $O(N^{d-1})$ time (without increasing the overall runtime).

\subsection{Extreme separator} \label{subsec2.1.1}

To solve the SP problem, we define a very important concept called \textit{extreme separator} through a sequence of steps.
Suppose a separable bichromatic dataset $T = T_R \cup T_B$ with SGPP is given in $\mathbb{R}^d$ ($d \geq 2$).
Assume that both $T_R$ and $T_B$ are nonempty.
Let $\mathcal{V}$ be the collection of the $(d-1)$-dim linear subspaces of $\mathbb{R}^d$ whose equation is of the form $ax_1+bx_2=0$,
where $a$ and $b$ are constants not equal to 0 simultaneously.
In other words, $\mathcal{V}$ contains all the $(d-1)$-dim linear subspaces that are perpendicular to the $x_1x_2$-plane and go through the origin.
Then there is a natural one-to-one correspondence between $\mathcal{V}$ and $\mathbb{P}^1$ (the 1-dim projective space),
\begin{equation*}
\sigma:[ax_1+bx_2=0]\  \longleftrightarrow\  [a:b].
\end{equation*}
For convenience, we use $\sigma$ to denote the maps in both directions in the rest of this paper. We now define a map $\pi_T:\mathcal{V} \rightarrow \{0,1\}$ as follows.
For any $V \in \mathcal{V}$, we orthogonally project all the points in $T$ onto $V$ and use $T^V = T_R^V \cup T_B^V$ to denote the new dataset after projection.
If $T^V$ is strongly separable, we set $\pi_T(V)=1$.
Otherwise, we set $\pi_T(V)=0$.
The map $\pi_T$ induces another map $\pi_T^*:\mathbb{P}^1 \rightarrow \{0,1\}$ by composing with the correspondence $\sigma$.
Let $P_0$ and $P_1$ be the pre-images of $\{0\}$ and $\{1\}$ under $\pi_T^*$, respectively (see Figure~\ref{fig:p0p1}).
By applying Theorem \ref{th1}, it is easy to prove the following.
\begin{theorem} \label{th2}
	$P_0$ is a connected closed subspace of $\mathbb{P}^1$.
	$P_0 = \emptyset$ iff $\Phi_J(T)$ is strongly separable in $\mathbb{R}^{d-2}$ for $J = \{3,4,\dots,d\}$.
\end{theorem}

We now have two cases, i.e., $P_0 \neq \emptyset$ and $P_0 = \emptyset$. If $P_0 \neq \emptyset$, we define the extreme separator of $T$ as follows.
Since $P_0$ is a connected closed subspace of $\mathbb{P}^1$, it has a unique clockwise boundary point $u^*$ (i.e., $u^*$ is the last point of $P_0$ in the clockwise direction).
Let $U^*=\sigma(u^*)$ be the linear subspace in $\mathcal{V}$ corresponding to $u^*$ (see Figure~\ref{fig:p0p1} again).
The following theorem reveals the separability property of $T^{U^*}$.
\begin{theorem} \label{th3}
	$T^{U^*}$ is weakly separable and there only exists one weak separator. 
	Furthermore, the unique separator of $T^{U^*}$ goes through exactly $d$ points, of which at least one is in $T_R^{U^*}$ and one is in $T_B^{U^*}$.
\end{theorem}

\begin{definition}
	(\textit{derived separator})
	Let $U$ be a $k$-dim linear subspace ($k<d$) of $\mathbb{R}^d$.
	Suppose $h$ is a strong (resp., weak) separator of $T^U$ in the space $U$.
	It is easy to see that the pre-image, $h'$, of $h$ under the orthogonal projection $\mathbb{R}^d \rightarrow U$ is a strong (resp., weak) separator of $T$ in $\mathbb{R}^d$.
	We call $h'$ the \textit{derived separator} of $h$ in $\mathbb{R}^d$.
\end{definition}

Let $h^*$ be the unique weak separator of $T^{U^*}$.
We define the \textit{extreme separator} of $T$ as the derived separator of $h^*$ in $\mathbb{R}^d$. (See Figure~\ref{fig:extreme}.)
At the same time, we call $U^*$ the \textit{auxiliary subspace} defining the extreme separator.
Clearly, the extreme separator and the auxiliary subspace are perpendicular to each other. 

On the other hand, if $P_0 = \emptyset$, we recursively define the extreme separator of $T$ as the derived separator of the extreme separator of $\Phi_J(T)$, for $J = \{3,4,\dots,d\}$.
Note that $P_0$ is nonempty when $d=2$.
To complete this recursive definition, we define the extreme separator in $\mathbb{R}^1$ as the weak separator (which is a point) with the smallest coordinate.

Note that the above definition of the extreme separator is only for the case that both $T_R$ and $T_B$ are nonempty.
In the trivial case where $T_R$ and/or $T_B$ is empty, we simply define the extreme separator as $x_d = \infty$.

To understand the intuition for the extreme separator, let us consider the case $d=3$.
Imagine there is a plane rotating clockwise around the $x_3$-axis.
We keep projecting the points in $T$ (orthogonally) to that plane and track the separability of the projection images.
If the images are always separable, then the extreme separator is defined recursively.
Otherwise, there is a closed period of time in which the images are inseparable, which is subsequently followed by an open period in which the images are separable.
At the connection of the two periods (from the inseparable one to the separable one), the images are weakly separable by a unique weak separator.
Then the rotating plane at this point is just the auxiliary subspace, and the extreme separator is obtained by orthogonally ``extending'' the unique weak separator to $\mathbb{R}^3$.

\subsection{Computing the separable-probability} \label{subsec2.1.2}
We remind the reader that $n = \min\{|S_R|,|S_B|\}$ and $N = \max\{|S_R|,|S_B|\}$.
Set $J = \{3,4,\dots,d\}$.
If the existent points in $S$ are separable, then there are two cases:
1) the extreme separator of the existent points is defined recursively (the case of $P_0 = \emptyset$) or equal to $x_d = \infty$ (the trivial case);
2) the extreme separator of the existent points is directly defined in $\mathbb{R}^d$ (the case of $P_0 \neq \emptyset$).
These two cases are clearly disjoint so that the SP can be computed as the sum of their probabilities.
By applying Theorem \ref{th2}, the probability of the first case is equal to the SP of $\Phi_J(S)$.
In the second case, according to Theorem \ref{th3}, the extreme separator goes through exactly $d$ points (of which at least one is red and one is blue).
Thus, the SP of $S$ can be computed as
\begin{equation*}
\mathit{Sep}(S) = \mathit{Sep}(\Phi_J(S)) + \sum_{h \in H_S} \tau_S(h),
\end{equation*}
where $H_S$ is the set of the hyperplanes that go through exactly $d$ points (of which at least one is red and one is blue) in $S$ and, for $h \in H_S$, $\tau_S(h)$ is the probability that the extreme separator of the existent points is $h$.

\begin{figure}[htpb]
    \centering
    \includegraphics{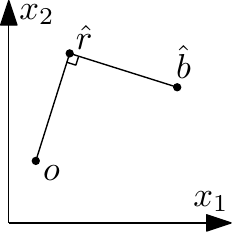}
    \caption{Illustrating the location of $o$. The space in the figure is the 2-dim subspace of $\mathbb{R}^d$ that is parallel to the $x_1x_2$-plane and contains $\hat{r}$, $\hat{b}$.}
    \label{fig:o}
\end{figure}


Clearly, for each $h \in H_S$, there is a unique element $U^* \in \mathcal{V}$ perpendicular to it ($h$ can never be parallel to the $x_1x_2$-plane due to the SGPP of $S$).
If $h$ is indeed the extreme separator of the existent points, then $U^*$ must be the auxiliary subspace.
Let $E = E_R \cup E_B$ be the set of the points on $h$.
In order to compute $\tau_S(h)$, we investigate the conditions for $h$ to be the extreme separator of the existent points.
First, as the $d$ points on $h$, the points in $E$ must exist.
Second, because the existent points should be \textit{weakly} (but not strongly) separable after being projected onto $U^*$,
there must exist $\hat{r} \in \mathcal{CH}(E_R)$ and $\hat{b} \in \mathcal{CH}(E_B)$ whose projection images on $U^*$ coincide, according to Theorem \ref{th1}. (Actually, such $\hat{r}$ and $\hat{b}$ are unique if they exist, due to the SGPP of $S$.)
Finally, since the extreme separator should weakly separate the existent points, all the red existent points must lie on one side of $h$ while all the blue ones must lie on the other side, except the points in $E$.
Also, the side for red/blue points is specific, as $\sigma(U^*)$ must be the \textit{clockwise} boundary of $P_0$.
To distinguish the red/blue side of $h$, we define, based on $\hat{r}$ and $\hat{b}$, an indicator $o = (o^{(1)},o^{(2)},\dots,o^{(d)})$, where \\
$
\begin{array}{l}
\rule{7mm}{0mm} o^{(1)} = \hat{r}^{(1)}+(\hat{b}^{(2)}-\hat{r}^{(2)}), \\
\rule{7mm}{0mm} o^{(2)} = \hat{r}^{(2)}+(\hat{r}^{(1)}-\hat{b}^{(1)}), \\
\rule{7mm}{0mm} o^{(i)} = \hat{r}^{(i)} = \hat{b}^{(i)}$ for $i \in J.
\end{array}
$\\
(See Figure~\ref{fig:o} for the location of $o$.)
It is easy to see that, when all the red (resp., blue) points appear on the same (resp., opposite) side of $h$ w.r.t. $o$, $\sigma(U^*)$ is the \textit{clockwise} boundary of $P_0$.
In sum, $h$ is the extreme separator of the existent points iff\\
\textbf{1)} all the points in $E$ exist; \\
\textbf{2)} there are $\hat{r} \in \mathcal{CH}(E_R)$ and $\hat{b} \in \mathcal{CH}(E_B)$ such that their projection images on $U^*$ coincide; \\
\textbf{3)} no red (resp. blue) point on the opposite (resp. same) side of $h$ w.r.t. $o$ exists.

Among the three conditions, the second one has nothing to do with the existences of the stochastic points in $S$ and can be verified in constant time.
If $h$ violates this condition, then $\tau_S(h)=0$.
Otherwise, $\tau_S(h)$ is just equal to the product of
the existence probabilities of the points in $E$ and the non-existence probabilities of the points which should not exist due to the third condition.
The simplest way to compute it is to scan every point in $S$ once, which takes linear time. 
This then leads to an $O(nN^d)$ overall time for computing the SP of $S$, since $|H_S|$ is bounded by $O(nN^{d-1})$. 

To reduce the time complexity, we can apply the idea of \textit{radial-order sort} in \cite{agarwal2014convex}. 
Specifically, when enumerating the hyperplanes spanned by $d$ points, we first determine $(d-1)$ points and sort all the remaining points around the $(d-2)$-dim subspace spanned by the those $(d-1)$ points (similar to polar-angle sorting around a point in $\mathbb{R}^2$).
Then we consider the last point in that sorted order and maintain a sliding window on the sorted list to record the points on one side of the current hyperplane.
In this way, each $\tau_S(h)$ can be computed in amortized constant time by modifying the previous result computed.
The time complexity is then reduced to $O(nN^{d-1} \log N)$ if we use $O(N \log N)$ time to do sorting each time.
Inspired by \cite{martin2015seperability}, we can further eliminate the log factor by taking advantage of {\it duality} \cite{deBerg_CG_book} and {\it topological sweep} \cite{Edelsbrunner:1986:topological_sweep} techniques.
Thus, the time complexity is finally improved to $O(nN^{d-1})$ for any $d \geq 3$.
See Appendix~\ref{append:improving_prob} for details.
(In the special case of $d=2$, the runtime becomes $O(N^2)$ instead of $O(nN)$ by applying this improvement so that the final time bound is $O(\min\{nN \log N, N^2\})$.)

\subsection{Witness-based lower bound for computing separable-probability}
When solving the SP problem, the key idea of our algorithm is to group the probabilities of those separable instances which share the same extreme separator so that the SP can be efficiently computed by considering the extreme separators instead of single instances.
Actually, by extending and abstracting this idea, we are able to get a general framework for computing SP, which we call the \textit{witness-based framework}.
Let $S$ be the given stochastic dataset and $\mathcal{I}_S$ be the set of all the separable instances of $S$.
The witness-based framework for computing the SP of $S$ is the following.
(Here, $\mathcal{P}(\cdot)$ denotes the powerset function.)

\begin{enumerate} 
	\item Define a set $W = \{h_1,\dots,h_m\}$ of hyperplanes (called \textit{witness separators}) with specified weights $w_1,\dots,w_m$ and an implicitly specified witness rule $f:W \rightarrow \mathcal{P}(\mathcal{I}_S)$ such that
	\begin{itemize}
		\item the instances in $f(h_i)$ are (either strongly or weakly) separated by $h_i$;
		\item the witness probability (see Step \textbf{2} below) of each $h_i$ is efficiently computable;
		\item any instance $I \in \mathcal{I}_S$ satisfies 
		$\sum\limits_{\forall i (I \in f(h_i))} w_i = 1$.
	\end{itemize}
	We say the witness separator $h_i$ \textit{witnesses} the instances in $f(h_i)$.
	\item Compute \textbf{efficiently} the \textit{witness probability} of each $h_i \in W$, which is defined as 
	\begin{equation*}
	\mathit{witP}(h_i) = \sum_{I \in f(h_i)} \mathit{Pr}(I),
	\end{equation*}
	i.e., the sum of the probabilities of all the instances witnessed by $h_i$.
	\item Compute the SP of $S$ by linearly combining the witness probabilities with the specified weights, i.e.,
	\begin{equation*}
	\mathit{Sep}(S) = \sum_{i=1}^{m}{ \left(w_i \cdot \mathit{witP}(h_i)\right) } = \sum_{I \in \mathcal{I}_S} \mathit{Pr}(I).
	\end{equation*}
\end{enumerate}
Note that the witness-based framework is very general.
The ways of defining witness separators and specifying witness rules may vary a lot among different witness-based algorithms.
Our algorithm and the one introduced in \cite{martin2015seperability}, which are the only two known algorithms for computing SP at this time, both belong to the witness-based framework.
Similar frameworks are also used to solve other probability-computing problems.
For example, the two algorithms in \cite{agarwal2014convex} for computing convex-hull membership probability are both implemented by defining witness edges/facets and summing up the witness probabilities.
To our best knowledge, up to now, most probability-computing problems under unipoint/multipoint model are solved by applying ideas close to this framework.

Now we show that any SP computing algorithm following the witness-based framework takes at least $\Omega(nN^{d-1})$ time in the worst case, and thus our algorithm is optimal among this category of algorithms for any $d \geq 3$.
Clearly, the runtime of a witness-based algorithm is at least $|W| = m$, i.e., the number of the witness separators.
Then a question naturally arises: how many witness separators do we need for computing SP?
From the above framework, one restriction for $W$ is that each separable instance of $S$ must be witnessed by at least one witness separator $h_i \in W$, i.e., $\mathcal{I}_S = \bigcup_{i=1}^{m} f(h_i)$.
Otherwise, the probabilities of the unwitnessed instances in $\mathcal{I}_S$ will not be counted when computing the SP of $S$.
It then follows that each separable instance of $S$ must be separated by some $h_i \in W$.
We prove that, in the worst case, we always need $\Omega (nN^{d-1})$ hyperplanes to separate all the separable instances of $S$, which implies an $\Omega (nN^{d-1})$ lower bound on the runtime of any witness-based SP computing algorithm.
We say a hyperplane set $H$ \textit{covers} a a bichromatic dataset $T = T_R \cup T_B$ iff for any non-trivial separable subset $V \subseteq T$ (i.e., $V$ contains at least one red point and one blue point), there exists $h \in H$ that separates $V$.
The following theorem completes the discussion, and is also of independent interest.

\begin{theorem} \label{lower}
	For a bichromatic dataset $T$, define $\chi(T)$ to be the size of the smallest hyperplane set that covers $T$.
	Let $\mathcal{T}_{n,N}^d$ be the collection of all the bichromatic datasets in $\mathbb{R}^d$ containing $n$ red points and $N$ blue points ($n \leq N$).
	Define
	\begin{equation*}
	\varGamma_d(n,N) = \sup_{T \in \mathcal{T}_{n,N}^d} \chi(T).
	\end{equation*}
	Then for any constant $d$, we have $\varGamma_d(n,N) = \Omega(nN^{d-1})$.
\end{theorem}

\section{Expected separation-margin}\label{sec:separation-margin}
In this section, we discuss how to compute the expected separation-margin (ESM) of a stochastic dataset $S = S_R \cup S_B$.
Again, we only describe the details of our algorithm under the unipoint model.
The generalization to the multipoint model is straightforward and is discuss in Appendix~\ref{sec:multipoint}.
We assume that $S$ has (conventional) general position property.

\subsection{Definitions} \label{subsec2.2.1}
Let $T = T_R \cup T_B$ be a separable bichromatic dataset and $h$ be a separator.
We define the margin of $h$ w.r.t. $T$ as
$M_h(T) = \min_{a \in T} \mathit{dist}(a,h).$
The separator which maximizes the margin is called the \textit{maximum-margin separator} and the corresponding margin is called the \textit{separation-margin} of $T$, denoted by $\mathit{Mar}(T)$.
If $T$ is not separable or if $T_R = \emptyset$ or $T_B = \emptyset$, we define its separation-margin to be 0 for convenience.
The ESM of a stochastic dataset $S = S_R \cup S_B$ is the expectation of the separation-margin of the existent points.
\begin{theorem} \label{th4} 
	For any separable dataset $T = T_R \cup T_B$ with $T_R \neq \emptyset$ and $T_B \neq \emptyset$, the maximum-margin separator of $T$ is unique.
	Furthermore, for any closest pair $(r,b)$ where $r \in \mathcal{CH}(T_R)$ and $b \in \mathcal{CH}(T_B)$, the maximum-margin separator of $T$ is the bisector of the segment $\overline{rb}$.
\end{theorem}

\begin{figure}[htpb]
    \centering
    \includegraphics[height=3.5cm]{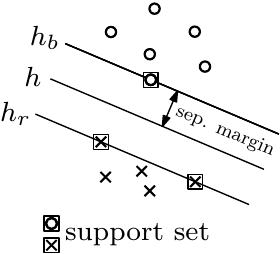}
    \caption{An example in $\mathbb{R}^2$}
    \label{fig:margin}
\end{figure}

Let $h$ be the maximum-margin separator of $T$ and $M = \mathit{Mar}(T)$ be its separation-margin.
Define $C_R = \{r \in T_R: \mathit{dist}(r,h) = M\}$ and $C_B = \{b \in T_B: \mathit{dist}(b,h) = M\}$.
We call $C = C_R \cup C_B$ the \textit{support set} of $T$ and the points in it the \textit{support points}.
All the support points have the same distance to the maximum-margin separator.
Thus, there should exist two parallel hyperplanes $h_r$ and $h_b$ (both parallel to the maximum-margin separator) where $h_r$ goes through all the red support points and $h_b$ goes through all the blue ones.
We call $h_r$ and $h_b$ the \textit{support planes} of $T$.
Including the maximum-margin separator $h$, they form a group of three parallel and equidistant hyperplanes $(h_r,h,h_b)$.
(See Figure~\ref{fig:margin})
Since the maximum-margin separator is unique, the support set and support planes are also unique.
We shall show that the maximum-margin separator can be uniquely determined via the support set.
\begin{theorem} \label{th5}
	Suppose $C$ is the support set of $T$.
	Then $T$ and $C$ share the same maximum-margin separator (also the same separation-margin) and the support set of $C$ is just itself.
\end{theorem}

\subsection{Computing the expected separation-margin} \label{subsec2.2.2}
According to Theorem \ref{th5}, the separation-margin of a separable dataset is equal to that of its support set.
Thus, the ESM of $S$ can be computed as
\begin{equation*}
\mathit{Emar}(S) = \sum_{C \subseteq S}{(\xi_S(C) \cdot \mathit{Mar}(C))},
\end{equation*}
where $\xi_S(C)$ is the probability that the existent points in $S$ are separable with the support set $C$.
Since $S$ has the general position property, the size of the support set of the existent points can be at most $2d$ ($d$ red points and $d$ blue points at most).
It follows that the total number of the possible $C$ to be considered is bounded by $O(n^d N^d)$.
Indeed, we can further improve this bound.
\begin{theorem} \label{th6}
	For a given stochastic dataset $S$ with general position property, the total number of the possible support sets is bounded by $O(nN^d)$.
	As a result, the number of the (distinct) possible separation-margins is also bounded by $O(nN^d)$.
\end{theorem}
By applying the previous formula for $\mathit{Emar}(S)$, we can enumerate all the $O(nN^d)$ possible support sets to compute the ESM of $S$.
The $O(nN^d)$ possible support sets can be enumerated as follows.
For the ones of sizes less than $(d+1)$, we enumerate them in the obvious way.
For the ones of sizes larger than or equal to $(d+1)$, we first enumerate a tuple of $(d+1)$ points (of which at least one is red and one is blue), which would be the representation of the support sets (see the proof of Theorem \ref{th6} in Appendix \ref{proof:th6}).
Via these $(d+1)$ points, we can determine two parallel hyperplanes $h_r$ and $h_b$ where $h_r$ goes through the red ones and $h_b$ goes through the blue ones.
We then find all the points on $h_r$ and $h_b$, the number of which is at most $2d$ (including the original $(d+1)$ points).
Once we have those points, we are able to enumerate all the possible support sets represented by the original $(d+1)$ points.
For each such possible support set $C$, $\mathit{Mar}(C)$ can be straightforwardly computed in constant time since $|C| \leq 2d$.
To compute $\xi_S(C)$, we observe that $C$ is the support set of the existent points iff\\
\textbf{1)} all the points in $C_R$ (resp., $C_B$) lie on $h_r$ (resp., $h_b$); \\
\textbf{2)} all the points in $C$ exist; \\
\textbf{3)} none of the red (resp., blue) points on the same side of $h_r$ (resp., $h_b$) w.r.t. $h$ exists; \\
\textbf{4)} except the points in $C$, none of the red (resp, blue) points on $h_r$ (resp., $h_b$) exists. \\
Among the four conditions, the first one has nothing to do with the existences of the stochastic points.
If the enumerated set, $C$, violates this condition, then $\xi_S(C)=0$.
Otherwise, $\xi_S(C)$ is just equal to the product of the existence probabilities of the points in $C$ (the second condition) and the non-existence probabilities of those points which should not exist (the last two conditions).
If we use the simplest way, i.e., scanning all the points in $S$, to find the points on $h_r$ and $h_b$ (for enumerating the possible support sets represented by a set of $(d+1)$ points) as well as to compute each $\xi_S(C)$, the total time for computing $\mathit{Emar}(S)$ is $O(nN^{d+1})$.
In fact, the runtime can be improved to $O(nN^d)$ by applying tricks similar to the ones used previously for improving the efficiency of our SP computing algorithm.
However, the way of applying them is somewhat different, and 
we present the details in Appendix~\ref{append:improving_margin}.

\subsection{Hardness of computing expected separation-margin}
We show that the bound achieved in Theorem~\ref{th6} is tight, which suggests that our algorithm for computing ESM may be difficult to improve further.
\begin{theorem} \label{tight}
	For any stochastic dataset $S$, define $\kappa(S)$ to be the total number of its (distinct) possible separation-margins.
	Then for any constant $d$, there exists some dataset $S$ containing $n$ red points and $N$ blue points ($n \leq N$) in $\mathbb{R}^d$ with $\kappa(S) = \Theta(n N^d)$.
\end{theorem}


From the above theorem, we can conclude that any algorithm that explicitly considers every possible separation-margin of the stochastic dataset requires at least $\Omega(n N^d)$ time to compute the ESM.
This then implies that our algorithm is optimal among this category of algorithms.
To do better, the only hope is to avoid considering every possible separation-margin explicitly.
However, this is fairly difficult (though may not be impossible) because of the lack of an explicit relationship among distinct separation-margins.

\section{Extension to general geometric objects} \label{sec:extension:shape}
\begin{figure}[htpb]
    \centering
    \includegraphics{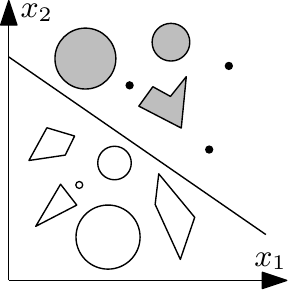}
    \caption{A separability problem for a set of bichromatic general objects in $\mathbb{R}^2$}
    \label{fig:objects}
\end{figure}

In the previous sections, we considered the separability related problems for stochastic points only.
In fact, the two problems can be naturally generalized to the case of general stochastic geometric objects (see Figure~\ref{fig:objects}).
In this paper, the general objects to be considered include polytopes with constant number of vertices, and/or $d$-dim closed balls with various radii.
We show that, with some effort, our methods can be extended to solve the generalized versions of the SP and ESM problems.
The stochastic model used is similar to the unipoint model: each object has a fixed location with an associated existence probability.
For convenience, we still use $S = S_R \cup S_B$ to denote the given stochastic dataset, in which each element is either a polytope or a ball.

\subsection{Reducing polytopes to points}
To deal with polytopes is easy, because of the fact that
the entire polytope is on one side of a (hyperplane) separator iff all its vertices are.
Thus, we can simply replace each polytope by its vertices and associate with each vertex an existence probability equal to that of the polytope.
In this way, the polytopes in $S$ can be reduced to points.
One thing should be noted is that, once we reduce the polytopes to points, the existences of the vertices of each polytope are dependent upon each other.
However, this issue can be easily handled without any increase in time complexity, because each polytope only has a constant number of vertices.

\subsection{Handling balls}\label{sec:ball}
Once we are able to use the vertices to replace the polytopes, it suffices to consider the separability problems for datasets containing only stochastic balls (points can be regarded as 0-radius balls).
Before we discuss how to handle balls, we need a definition of general position for a ball-dataset.
We say a set of balls in $\mathbb{R}^d$ is in general position (or has the general position property) if \\
\textbf{1)} the centers of the balls are in general position; \\
\textbf{2)} no $(d+1)$ balls have a common tangent hyperplane. \\
Furthermore, we say a ball-dataset has strong general position property (SGPP) if it satisfies the two conditions above and all of the 0-radius balls in it have SGPP (as defined in Section~\ref{sec:separable-prob}) when regarded as points.
In Section~\ref{sp-ball}, the given ball-dataset $S$ is required to have SGPP.
And in Section~\ref{esm-ball}, we only need the assumption that $S$ has the (usual) general position property.

\subsubsection{Separable-probability (ball-version)} \label{sp-ball}
Let $T=T_R \cup T_B$ be a set of bichromatic balls with SGPP and set $J = \{3,4,\dots,d\}$.
With similar proofs, Theorem~\ref{th1} and \ref{th2} can be directly generalized to ball-datasets
(the meaning of $\mathcal{CH}(T_R)$/$\mathcal{CH}(T_B)$ should be modified as the convex hull of all the balls in $T_R$/$T_B$).
The ball-version of Theorem \ref{th3} (and also its proof) is slightly different, which we present as follows (the proof can be found in Appendix~\ref{th3ball-proof}).
\begin{theorem} \label{th3ball}
	$T^{U^*}$ is weakly separable and there only exists one weak separator.
	Furthermore, the unique weak separator of $T^{U^*}$ either goes through exactly $d$ 0-radius balls (of which at least one is in $T_R^{U^*}$ and one is $T_B^{U^*}$) or is tangent to at least one ball with radius larger than 0.
\end{theorem}
Once we generalize those results, we are immediately able to generalize the concept of extreme separator to ball-datasets.
As we do in Section~\ref{subsec2.1.1}, if $P_0 \neq \emptyset$, we define the extreme separator of $T$ as the derived separator of the unique weak separator of $T^{U^*}$.
If $P_0 = \emptyset$, we recursively define the extreme separator of $T$ as the derived separator of the extreme separator of $\Phi_J(T)$.
If the extreme separator is directly defined (i.e., the case of $P_0 \neq \emptyset$), we call the subset consisting of all the balls tangent to extreme separator the \textit{critical set}.
We shall use the following theorem later for solving the ball version of the SP problem.
\begin{theorem} \label{th7}
	Let $T=T_R \cup T_B$ be a separable bichromatic ball-dataset whose extreme separator is directly defined and let $C$ be its critical set.
	Then the extreme separator of $C$ is also directly defined.
	Furthermore, $T$ and $C$ share the same extreme separator and auxiliary subspace.
\end{theorem}

Theorem~\ref{th7} implies that the extreme separator is uniquely determined by the critical set.
This then gives us the basic idea to solve the problem, enumerating all the possibilities for the critical set.
As in Section~\ref{subsec2.1.2}, we can compute the SP of $S$ as
\begin{equation*}
\mathit{Sep}(S) = \mathit{Sep}(\Phi_J(S)) + \sum_{C \subseteq S} \lambda_S(C),
\end{equation*}
where $\lambda_S(C)$ is the probability that the critical set of the existent balls is $C$.
Since the balls in $S$ have SGPP, the size of the critical set of the existent balls can be at most $d$.
Furthermore, the critical set should contain at least one red ball and one blue ball.
Thus, it suffices to compute $\lambda_S(C)$ for all the subsets $C \subseteq S$ with $|C| \leq d$ which contain balls of both colors.
We consider two cases separately.
First, all the balls in $C$ have radius 0.
Second, there is at least one ball in $C$ with radius larger than 0. 

In the first case, according to Theorem~\ref{th3ball}, $\lambda_S(C)>0$ only if $|C|=d$.
Since the balls in $C$ are actually points, the situation here is almost the same as what we confronted in the point-version of the problem.
We can uniquely determine a hyperplane $h$ which goes through the $d$ points in $C$, and a subspace $U^* \in \mathcal{V}$ perpendicular to $h$.
Then $\lambda_S(C)$ is just equal to the probability that $h$ is the extreme separator of the existent balls.
Also, the conditions for $h$ to be the extreme separator are very similar to those in Section~\ref{subsec2.1.2}, which are \\
\textbf{1)} all the balls in $C$ exist; \\
\textbf{2)} there exist $r \in \mathcal{CH}(C_R)$ and $b \in \mathcal{CH}(C_B)$ such that their projection images on $U^*$ coincide; \\
\textbf{3)} no red (resp. blue) ball on the opposite (resp. same) side of $h$ w.r.t. the point $o$ exists,
where the definition of $o$ is similar to that in Section~\ref{subsec2.1.2}; \\
\textbf{4)} no ball intersecting with $h$ exists, except the ones in $C$. \\
If $C$ violates the second condition, then $\lambda_S(C) = 0$.
Otherwise, $\lambda_S(C)$ is just equal to the product of
the existence probabilities of the balls in $C$
and the non-existence probabilities of the balls that should not exist.

In the second case, however, the size of $C$ may be less than $d$.
According to Theorem~\ref{th7}, if $C$ is the critical set of the existent points, then the extreme separator and auxiliary subspace of the existent points are the same as those of $C$.
This implies that $\lambda_S(C)=0$ if $C$ is not separable or the extreme separator of $C$ is defined recursively.
So we only need to consider the situation that the extreme separator of $C$ is directly defined.
Assume that $C$ has the extreme separator $h$ (directly defined) with the auxiliary subspace $U^* \in \mathcal{V}$.
Let $c$ be any ball in $C$ with radius larger than 0.
Then it is easy to see that $C$ is the critical set of the existent balls iff \\
\textbf{1)} all the balls in $C$ exist; \\
\textbf{2)} all the balls in $C$ are tangent to $h$; \\
\textbf{3)} no ball with the same color as (resp. different color than) $c$ but on the opposite (resp. same) side of $h^*$ w.r.t. $c$ exists; \\
\textbf{4)} no ball intersecting with $h$ exists, except the ones in $C$. \\
Because of the constant size of $C$, $h$ and $U^*$ can be computed in constant time by simply using brute-force.
Similarly, if $C$ satisfies the second condition, $\lambda_S(C)$ is equal to the product of
the existence probabilities of the balls in $C$
and the non-existence probabilities of the balls that should not exist. 

In both the cases, $\lambda_S(C)$ can be computed in linear time by simply scanning all the balls in $S$.
Thus, the SP can be finally computed in $O(nN^d)$ time, as the number of the subsets $C$ considered is bounded by $O(nN^{d-1})$.
Unfortunately, the improvement techniques used in the point-version of the problem cannot be generalized to ball-datasets so that our eventual time bound for computing the SP of general geometric objects remains $O(nN^d)$.

\subsubsection{Expected separation-margin (ball-version)} \label{esm-ball}
Let $T = T_R \cup T_B$ be a set of bichromatic balls in general position.
Clearly, the definitions given in Section~\ref{subsec2.2.1} (maximum-margin separator, separation-margin, support set/points/planes, etc.) can be directly generalized to ball-datasets.
Also, with these definitions, the ball-versions of Theorem \ref{th4} and \ref{th5} can be easily verified (by using the same proofs).

To extend the previous algorithm, we need to prove the ball version of Theorem \ref{th6}.
The first step is the same as that in the original proof of Theorem \ref{th6}: we arbitrarily label the balls in $S$ and define the representation of $C$ as the $(d+1)$ balls in $C$ with the smallest labels, for any $C \subseteq S$ with $|C|>d$.
We show that the number of possible support sets represented by any group of $(d+1)$ balls is $O(1)$.
Let $a_1,a_2,\dots,a_{d+1}$ be any $(d+1)$ balls in $S$ where $a_1,\dots,a_k$ are red and $a_{k+1},\dots,a_{d+1}$ are blue,
where $1 \leq k \leq d$ as before.
Let each ball $a_i$ have center $c_i$ and radius $\delta_i$.
If some possible support set $C$ is represented by these $(d+1)$ balls, then the support plane $h_r$ (resp. $h_b$) must be tangent to $a_1,\dots,a_k$ (resp. $a_{k+1},\dots,a_{d+1}$).
Furthermore, the balls $a_1,\dots,a_k$ (resp. $a_{k+1},\dots,a_{d+1}$) must be on the open side of $h_r$ (resp. $h_b$), i.e.,
the side different from the one containing the area between $h_r$ and $h_b$.
Formally, suppose the equations of $h_r$ and $h_b$ are $\vec{\omega} \cdot x + b_1 = 0$ and $\vec{\omega} \cdot x + b_2 = 0$.
We then have the following system of equations 
\begin{equation*}
\left\{
\begin{array}{cl}
\vec{\omega} \cdot c_i + b_1 = -r_i & \text{for } i \in \{1,\dots,k\}, \\
\vec{\omega} \cdot c_i + b_2 = r_i & \text{for } i \in \{(k+1),\dots,(d+1)\}, \\
| \vec{\omega} | = 1, \\
b_1<b_2.
\end{array}
\right.
\end{equation*}
The $(d+1)$ linear equations are linearly independent, as the centers are in general position.
Thus, by limiting the norm of $\vec{\omega}$ to be 1, this system has at most two solutions.
In other words, there are at most two possibilities for the support planes $(h_r,h_b)$.
By following the logic in the proof of Theorem \ref{th6}, we then know the number of the possible support sets represented by these $(d+1)$ balls is constant,
which immediately implies that the total number of all possible support sets is bounded by $O(nN^d)$.

To enumerate these possible support sets, we can directly use the same method as in Section~\ref{subsec2.2.2}, i.e., 
first enumerate $(d+1)$ balls and then enumerate the possible support sets represented by them.
Again, because the improvement techniques used in the point-version of the problem do not work for ball-datasets,
we have to scan all the balls once for computing the corresponding probability of each possible support set,
which makes the overall time $O(nN^{d+1})$ for computing the ESM of general geometric objects.

\section{Applications}\label{sec:application}

In this section, we present some applications of our algorithms to stochastic convex-hull (SCH) related problems.
Given a stochastic (point) dataset $A$, the SCH of $A$ refers to the convex-hull of the existent points in $A$, which is an uncertain convex shape.

\subsection{SCH membership probability problem}
The SCH membership probability problem was introduced for the first time in \cite{agarwal2014convex}.
The problem can be described as follows:
given a stochastic dataset $A = \{a_1,\dots,a_m\} \subset \mathbb{R}^d$ and a query point $q \in \mathbb{R}^d$, compute the probability that $q$ is inside the SCH of $A$, which we call the \textit{SCH membership probability} (SCHMP) of $q$ w.r.t. $A$.

It is shown in \cite{martin2015seperability} that the SCHMP problem in $\mathbb{R}^d$ can be reduced to the SP problem in $\mathbb{R}^{d-1}$.
Due to this, by plugging in our SP computing algorithm presented in Section \ref{sec:separable-prob}, we immediately obtain an $O(m^{d-1})$-time algorithm to compute SCHMP for $d \geq 3$, which matches the best known bound in \cite{martin2015seperability}.
Indeed, this bound can be achieved by applying any SP computing algorithm with runtime bounded by $O(N^d)$.

More interestingly, we show that our SP computing algorithm yields a more direct and natural method to solve the SCHMP problem in $O(m^{d-1})$ time for $d \geq 3$ and $O(m \log m)$ time for $d = 2$, which does not involve any non-trivial reduction between the two problems.
Given a SCHMP problem instance $(A,q)$, clearly, the query point $q$ is outside the SCH of $A$ iff it can be separated from the existent points in $A$ by a hyperplane.
Thus, we construct a stochastic bichromatic dataset $S = S_R \cup S_B$, where $S_R$ contains only one point, $q$, with existence probability 1 and $S_B = A$.
Then the SCHMP of $q$ w.r.t. A is just equal to $1-\mathit{Sep}(S)$.
This can be computed in $O(m^{d-1})$ time for $d \geq 3$ and $O(m \log m)$ time for $d = 2$ by applying our SP computing algorithm, since $|S_R| = 1$ and $|S_B| = m$.
Note that the $O(m^{d-1})$ runtime of this simple method relies on the $O(nN^{d-1})$ time bound of our SP computing algorithm for $d \geq 3$.
If we plug in an $O(N^d)$-time SP computing algorithm, the time cost will become $O(m^d)$.
Interestingly enough, this method for computing SCHMP is a generalization of the witness-edge method in \cite{agarwal2014convex} to the case $d>2$, where the latter was the first known approach that solves this problem in $\mathbb{R}^2$ and was thought to be difficult to be generalized to higher dimensions \cite{agarwal2014convex}.
This can be seen as follows.
When plugging in our SP computing algorithm, we enumerate all the possible extreme separators of $\{q\} \cup \varGamma$, where $\varGamma$ denotes the set of the existent points in $A$.
If the extreme separator is finally defined in $\mathbb{R}^{d-2k}$, it goes through $(d-2k)$ points, of which one is $q$.
These $(d-2k)$ points form a $(d-2k-1)$-dim face of $\mathcal{CH}(\{q\} \cup \varGamma)$ about the vertex $q$.
It is evident that this face is uniquely determined by the convex polytope $\mathcal{CH}(\{q\} \cup \varGamma)$.
We call it the \textit{witness-face} of $q$ in $\mathcal{CH}(\{q\} \cup \varGamma)$.
Then enumerating the possible extreme separators is equivalent to enumerating the possible witness-faces of $q$ in $\mathcal{CH}(\{q\} \cup \varGamma)$.
When $d = 2$, the concept of witness-face coincides with that of \textit{witness-edge} defined in \cite{agarwal2014convex}.
Thus, in this case, our method is identical to the witness-edge method.

\subsection{Other SCH-related problems}
Our algorithms presented in the previous sections can also be applied to solve some new problems related to SCH.
Here we propose three such problems and show how to solve them. \\
$\bullet$ \textbf{SCH intersection probability problem}.
This problem is a natural generalization of the SCHMP problem.
Given a stochastic dataset $A = \{a_1,\dots,a_m\} \subset \mathbb{R}^d$ and a query object $Q$ which is a convex polytope with constant complexity (e.g., segment, simplex, etc.) in $\mathbb{R}^d$, the goal is to compute the probability that $Q$ has non-empty intersection with the SCH of $A$.
When $Q$ is a single point, this is just the SCHMP problem.
To solve this problem, we extend the method described in the preceding subsection.
According to Theorem \ref{th1}, $Q$ has no intersection with the SCH iff its vertices can be separated from the existent points in $A$ by a hyperplane.
Based on this, by constructing a stochastic bichromatic dataset $S = S_R \cup S_B$, where $S_R$ contains the vertices of $Q$ with existence probability 1 and $S_B = A$, we can apply our SP computing algorithm to compute the desired probability in $O(m^{d-1})$ time (note that $|S_R|$ is $O(1)$ for $Q$ has constant complexity). \\
$\bullet$ \textbf{SCH $\varepsilon$-distant probability problem}.
This problem is another natural generalization of the SCHMP problem.
Given a stochastic dataset $A = \{a_1,\dots,a_m\} \subset \mathbb{R}^d$, a query point $q \in \mathbb{R}^d$, and a parameter $\varepsilon \geq 0$, the goal is to compute the probability that the distance from $q$ to the SCH of $A$ is greater than $\varepsilon$.
When $\varepsilon = 0$, this is equivalent to the SCHMP problem.
To solve this problem, we need to apply our generalized SP computing algorithm presented in Section \ref{sec:extension:shape}.
Clearly, $q$ has a distance greater than $\varepsilon$ to the SCH of $A$ iff the $\varepsilon$-ball centered at $q$ can be separated from the existent points in $A$ by a hyperplane.
Thus, by constructing a generalized stochastic bichromatic dataset $S = S_R \cup S_B$, where $S_R$ contains the $\varepsilon$-ball centered at $q$ with existence probability 1 and $S_B = A$, we can apply our generalized SP computing algorithm to compute the desired probability in $O(m^d)$ time. \\
$\bullet$ \textbf{Expected distance to a SCH}.
Given a stochastic dataset $A = \{a_1,\dots,a_m\} \subset \mathbb{R}^d$ and a query point $q \in \mathbb{R}^d$, the goal of this problem is to compute the expected distance from $q$ to the SCH of $A$.
To achieve this, we notice that the distance from $q$ to the SCH of $A$ is just equal to the separation-margin of $\{q\} \cup \varGamma$, where $\varGamma$ denotes the set of the existent points in $A$.
Thus, we construct a stochastic bichromatic dataset $S = S_R \cup S_B$, where $S_R$ contains only one point $q$ with existence probability 1 and $S_B = A$.
Then the problem can be solved in $O(m^d)$ time by plugging in our ESM computing algorithm presented in Section \ref{sec:separation-margin}.

\bibliography{lipics-v2016-jie-article}

\newpage
\appendix
\section{Detailed proofs} \label{append.proof}
\subsection{Proof of Theorem \ref{th1}}
The ``only if'' part is obvious. 
Suppose we have a hyperplane $h$ which separates $T_R$ and $T_B$ into different sides. 
Let $H$ be the half-space which contains $T_R$ and $H'$ be the other one which contains $T_B$. 
Since both $H$ and $H'$ are convex, we have $\mathcal{CH}(T_R) \subseteq H$ and $\mathcal{CH}(T_B) \subseteq H'$. 
It immediately follows $\mathcal{CH}(T_R) \cap \mathcal{CH}(T_B) = \emptyset$. 
To prove the ``if'' part, we assume $\mathcal{CH}(T_R) \cap \mathcal{CH}(T_B) = \emptyset$. 
Let $(r,b)$ be the closest point-pair where $r \in \mathcal{CH}(T_R)$ and $b \in \mathcal{CH}(T_B)$. 
We denote the mid-point of the segment $\overline{rb}$ by $s$ and define a separator $h$ as the hyperplane that goes through $s$ and perpendicular to $\overline{rb}$. 
We prove that $h$ separates $T_R$ and $T_B$ by contradiction. 
Assume $h$ does not separate $T_R$ and $T_B$. 
That means there are two points in $T_R$ (or $T_B$) on the different sides of $h$. 
Without loss of generality, we just assume such two points are in $T_R$. 
Thus, we can find a point $r^* \in \mathcal{CH}(T_R)$ which is on the hyperplane $h$. 
If we observe the triangle $\triangle brr^*$, we find $\angle brr^* < \pi/2$,
which means there exists a point $t \in \overline{rr^*}$ such that $\mathit{dist}(t,b)<\mathit{dist}(r,b)$. 
This contradicts the fact that $(r,b)$ is the closest point-pair (note that $t \in \mathcal{CH}(T_R)$). 
Thus, $h$ separates $T_R$ and $T_B$.

\subsection{Proof of Theorem \ref{th2}} \label{th2-proof}
We define a subset of $\mathcal{CH}(T_R) \times \mathcal{CH}(T_B)$ as
\begin{equation*}
D = \{(r,b) \in \mathcal{CH}(T_R) \times \mathcal{CH}(T_B):\phi_J(r)=\phi_J(b)\}
\end{equation*}
where $J = \{3,4,\dots,d\}$.
Also, we define a continuous function $f:D \rightarrow \mathbb{P}^1$ as
\begin{equation*}
f: (r,b) \longmapsto [(r^{(1)}-b^{(1)}):(r^{(2)}-b^{(2)})].
\end{equation*}
We shall first prove that $P_0$ is equal to the image of $f$, $\text{Im}f$.
Let $u = [a:b]$ be a point in $\mathbb{P}^1$ and $U = \sigma(u)$.
According to Theorem \ref{th1}, $u \in P_0$ iff $\mathcal{CH}(T_R^U) \cap \mathcal{CH}(T_B^U) \neq \emptyset$.
It is clear that $\mathcal{CH}(T_R^U) \cap \mathcal{CH}(T_B^U) \neq \emptyset$ iff $u$ is in the image of $f$, which implies $P_0 = \text{Im}f$.
Then it suffices to prove the theorem regarding $\text{Im}f$ instead of $P_0$.
Because of the connectedness and compactness of $D$, $\text{Im}f$ is also connected and compact.
Furthermore, since $\mathbb{P}^1$ is Hausdorff, $\text{Im}f$ is closed.
Thus, the first statement of the theorem is proved.
To prove the second one, we first assume $\text{Im}f = \emptyset$, which implies $D = \emptyset$.
It then immediately follows that $\Phi_J(T)$ is strongly separable in $\mathbb{R}^{d-2}$ for $J = \{3,4,\dots,d\}$.
On the other hand, if $\Phi_J(T)$ is strongly separable, $\mathcal{CH}(\Phi_J(T_R)) \cap \mathcal{CH}(\Phi_J(T_B)) = \emptyset$.
In this situation, $D$ has to be empty and thus $\text{Im}f = \emptyset$.

\subsection{Proof of Theorem \ref{th3}}
Suppose that $\alpha_{\vec{v}} = \min_{r \in T_R}\{\vec{v} \cdot r\}$, $\alpha'_{\vec{v}} = \max_{r \in T_R}\{\vec{v} \cdot r\}$, $\beta_{\vec{v}} = \min_{b \in T_B}\{\vec{v} \cdot b\}$, $\beta'_{\vec{v}} = \max_{b \in T_B}\{\vec{v} \cdot b\}$.
We define a function $f:\mathbb{P}^1 \rightarrow \mathbb{R}$ as
\begin{equation*}
f(u) = \sup_{\vec{v} \in \bar{U}}\  \max\{(\alpha_{\vec{v}}-\beta'_{\vec{v}}),(\beta_{\vec{v}}-\alpha'_{\vec{v}})\},
\end{equation*}
where $\bar{U} = \mathbb{S}^{d-1} \cap \sigma(u)$ ($\mathbb{S}^{d-1}$ is the unit sphere in $\mathbb{R}^d$).
It is easy to see that $f$ is continuous.
Furthermore, according to the definition of $P_0$, we know that $u \in P_0$ iff $f(u) \leq 0$.
Since $u^*$ is a boundary point of $P_0$, we have $f(u^*)=0$.
Thus, $T^{U^*}$ is weakly separable.
To prove the remaining part of the theorem, we introduce a definition called \textit{degree}.
Let $X$ be a polytope and $x$ be a point on the boundary of $X$.
We define the \textit{degree} of $x$ in $X$, denoted by $\deg_Xx$, to be the minimum of the dimensions of all the simplices that are spanned by some vertices of $X$ and contain $x$.
Since $T^{U^*}$ is not strongly separable, we can find a point $x^* \in \mathcal{CH}(T_R^{U^*}) \cap \mathcal{CH}(T_B^{U^*})$.
To simplify the notation, we denote $\mathcal{CH}(T_R^{U^*})$ by $C_1$ and $\mathcal{CH}(T_B^{U^*})$ by $C_2$.
We claim that $\deg_{C_1}x^* + \deg_{C_2}x^* \geq d-2$.
Suppose $\deg_{C_1}x^* = k_1$ and $\deg_{C_2}x^* = k_2$.
According to the definition of degree, we can find $(k_1+1)$ red (resp. $(k_2+1)$ blue) points in $T_R^{U^*}$ (resp. $T_B^{U^*}$)
such that the simplex spanned by these points, $\bar{s}_R$ (resp. $\bar{s}_B$), contains $x^*$ in its interior.
Let $\alpha:\mathbb{R}^d \rightarrow U^*$ and $\beta:U^* \rightarrow \mathbb{R}^{d-2}$ be the orthogonal projection functions.
Clearly, we have
\begin{equation*}
\phi_J = \beta \circ \alpha,
\end{equation*}
for $J = \{3,4,\dots,d\}$.
Then the convex hull of the $\beta$-images of the vertices of $\bar{s}_R$ (resp. $\bar{s}_B$) contains the point $\beta(x^*)$.
The $\beta$-images of the points in $T^{U^*}$ are just the points in $\Phi_J(T)$.
If $k_1 + k_2 < d-2$, we can always find two simplices with the vertices in $\Phi_J(T)$ such that they intersect at $\beta(x^*)$ and the sum of their dimensions is less than $(d-2)$.
This contradicts the fact that $\Phi_J(T)$ is in general position (as $T$ has SGPP).
Thus, $\deg_{C_1}x^* + \deg_{C_2}x^* \geq d-2$.
Now we go back to $U^*$.
We know that $T^{U^*}$ is weakly separable.
Let $h$ be a weak separator.
Since $x^* \in C_1 \cap C_2$, $x^*$ must be on $h$.
Note that $x^*$ is in the interiors of $\bar{s}_R$ and $\bar{s}_B$.
This implies $h$ must go through all of the $(k_1+k_2+2)$ vertices of $\bar{s}_R$ and $\bar{s}_B$.
Since $k_1+k_2+2 \geq d$ and $T$ is in general position, the weak separator $h$ is unique and goes through exactly $d$ points in $T^{U^*}$ (of which at least one is in $T_R^{U^*}$ and one is $T_B^{U^*}$).

\subsection{Proof of Theorem \ref{lower}}
To prove this theorem, it is more convenient to work on ``directed'' hyperplanes.
A \textit{directed hyperplane} in $\mathbb{R}^d$ is a hyperplane with one side (half-space) specified to be red and the other side specified to be blue.
It can be represented as a $(d+1)$-tuple $(a_0,a_1,\dots,a_d)$ of real numbers (not all equal to 0 simultaneously) such that the inequality $a_0 + \sum_{i=1}^{d} a_i x_i < 0$ indicates the red side.
We say the directed hyperplane $(a_0,a_1,\dots,a_d)$ \textit{separates} a set of bichromatic points iff there is no point located on the side of different color, i.e., for each point $x = (x_1,\dots,x_d)$, we have
\begin{equation*}
a_0 + \sum_{i=1}^{d} a_i x_i \left\{
\begin{array}{ll}
\leq 0 & \text{if } x \text{ is red}, \\
\geq 0 & \text{if } x \text{ is blue}.
\end{array}
\right.
\end{equation*}
Since a (undirected) hyperplane can be replaced with two directed hyperplanes, the number of the directed hyperplanes required for covering a dataset is at most twice the number of the undirected ones.
Thus, it suffices to prove the result with respect to directed hyperplanes.
In the rest of the proof, the notation $\chi(T)$ is used to denote the size of the smallest directed-hyperplane set (instead of hyperplane set) which covers $T$.

We show that, for any constant $d$, there exists some bichromatic dataset $T \in \mathcal{T}_{n,N}^d$ with general position such that $\chi(T)=\Omega(nN^{d-1})$.
Specifically, we use induction on the dimension $d$.
The base case $d=1$ is trivial.
Assume that for any $d \leq k-1$, such bichromatic dataset $T$ exists.
Now we try to construct $T$ in $\mathbb{R}^k$.

Our first step is to construct a set $T'$ of one red point and $\Theta(N)$ blue points in $\mathbb{R}^k$ with $\chi(T') = \Omega (N^{k-1})$.
Let $U = U_R \cup U_B$ be a set of $N$ red points and $N$ blue points (in general position) in $\mathbb{R}^{k-1}$ with $\chi(U) = \Omega (N^{k-1})$.
Define two functions $f_R, f_B: \mathbb{R}^{k-1} \rightarrow \mathbb{R}^k$ as
\begin{equation*}
f_R: (x_1,\dots,x_{k-1}) \mapsto (-x_1,\dots,-x_{k-1},-1),
\end{equation*}
\begin{equation*}
f_B: (x_1,\dots,x_{k-1}) \mapsto (x_1,\dots,x_{k-1},1).
\end{equation*}
Then we set the $2N$ blue points in $T'$ to be $f_R(U_R) \cup f_B(U_B)$ (ignoring their original colors in $U$) and the only red point in $T'$ to be the origin of $\mathbb{R}^k$.
We claim that $\chi(T') = \Omega (\chi(U))$.
For any non-trivial separable subset $V \subseteq U$ (i.e., $V$ contains at least one red point and one blue point), define $f(V)$ to be a subset of $T'$ containing the blue points $f_R(V_R) \cup f_B(V_B)$ and the only red point.
It is easy to see that $V$ is separable iff $f(V)$ is.
Furthermore, if a non-horizontal (i.e., not parallel to the plane $x_d = 0$) directed hyperplane in $\mathbb{R}^k$, $(a_0,a_1,\dots,a_k)$, separates $f(V)$, then we have a corresponding directed hyperplane in $\mathbb{R}^{k-1}$, $(a_0+a_k,a_1,\dots,a_{k-1})$, which separates $V$.
We call the latter the \textit{induced} plane of the former.
Now let $H = \{h_1,\dots,h_{\chi(T')}\}$ be a set of directed hyperplanes in $\mathbb{R}^k$ which cover $T'$.
Assume they are all non-horizontal (if any of them is horizontal, we can always slightly rotate it without changing the subsets of $T'$ it separates).
Then let $H' = \{h'_1\dots,h'_{\chi(T')}\}$ be a set of directed hyperplanes in $\mathbb{R}^{k-1}$ in which $h'_i$ is the induced plane of $h_i$.
Clearly, $H'$ covers $U$, which implies that $\chi(U) \leq \chi(T')$.

The next step is to extend $T'$ into another set $T$ of $\Theta(n)$ red points and $\Theta(N)$ blue points in $\mathbb{R}^k$ with $\chi(T) = \Omega(nN^{k-1})$.
We denote by $r$ the only red point in $T'$ (whose location is the origin of $\mathbb{R}^k$) and by $b_1,\dots,b_{2N}$ the $2N$ blue points in $T'$.
We first slightly perturb each $b_i$ without changing $\chi(T')$ to make the points $r,b_1,\dots,b_{2N}$ be in general position.
For convenience, we now use $T'$ to denote the new set after the perturbation.
Then we find an $\varepsilon$-ball centered at the origin of $\mathbb{R}^k$ with $\varepsilon$ small enough such that if the red point $r$ perturbs inside that ball, $\chi(T')$ does not change.
The value of $\varepsilon$ can be determined as follows.
For each $(k-1)$-dim linear subspace spanned by $k$ blue points $b_{\pi_1},\dots,b_{\pi_k}$, we compute the distance from the origin to it.
And $\varepsilon$ is then set to be a number less than the minimum of those distances.
Inside this $\varepsilon$-ball, we pick $n$ red points $r_1,\dots,r_n$ such that all the points $r_1,\dots,r_n,b_1,\dots,b_{2N}$ are in general position.
Define the red points in $T$ to be $r_1,\dots,r_n$.
Next, we find another small number $\varepsilon'$ such that for any hyperplane $h$ in $\mathbb{R}^k$, there are at most $k$ points among $r_1,\dots,r_n$ whose distances to $h$ are at most $\varepsilon'$.
We can determine $\varepsilon'$ as follows.
For each $(k+1)$-tuple $t = (r_{\pi_1},\dots,r_{\pi_{k+1}})$, we define
\begin{equation*}
\delta_t = \inf_h \max_{i=1}^{k+1} \mathit{dist}(h,r_{\pi_i}).
\end{equation*}
Clearly, $\varepsilon'$ can be any number less than the minimum of all $\delta_t$.
Now, for each red point $r_i$, we add $(k+1)$ blue points $b'_{i,1},\dots,b'_{i,{k+1}}$ inside the $\varepsilon'$-ball centered at $r_i$ such that the simplex spanned by $b'_{i,1},\dots,b'_{i,{k+1}}$ contains $r_i$ in its interior.
We carefully determine the locations of these additional points to guarantee general position.
Define the blue points in $T$ to be these $(k+1)n$ additional points and the original $2N$ ones.
We prove that $\chi(T) = \Omega(nN^{k-1})$.
Let $H$ be any directed-hyperplane set which covers $T$.
Also, let $H_i \subseteq H$ be the subset of the directed hyperplanes whose distances to the point $r_i$ are at most $\varepsilon'$.
We claim that $|H_i| \geq \chi(T')$ for any $i \in \{1,\dots,n\}$.
Set $T'' = T''_R \cup T''_B = \{r_i\} \cup \{b_1,\dots,b_{2N}\}$.
Recall that $r_i$ is inside the $\varepsilon$-ball centered at the origin of $\mathbb{R}^k$, which implies $\chi(T'') = \chi(T')$.
Assume that $|H_i| < \chi(T'')$.
Then there exists a non-trivial separable subset $V \subseteq T''$ which is not separated by any $h \in H_i$.
Let $h^*$ be a directed hyperplane which goes through $r_i$ and weakly separates $V$.
Consider the blue points $b'_{i,1},\dots,b'_{i,k+1}$.
Since $r_i$ is in the interior of the simplex spanned by $b'_{i,1},\dots,b'_{i,k+1}$, we can find at least one point $b'_{i,j}$ such that the subset of $T$, $V \cup \{b'_{i,j}\}$, is also separated by $h^*$ (and thus separable).
We show that $V \cup \{b'_{i,j}\}$ is not separated by any $h \in H$, which contradicts the fact that $H$ covers $T$.
We consider two cases: $h \in H_i$ and $h \in H \backslash H_i$.
Any $h \in H_i$ is not a separator of $V \cup \{b'_{i,j}\}$ because it does not separate $V$.
For any $h \in H \backslash H_i$, we notice that $\mathit{dist}(h,r_i) > \epsilon'$.
Thus, both $r_i$ and $b'_{i,j}$ are on the same side of $h$, which implies that $h$ is not a separator of $V \cup \{b'_{i,j}\}$.
As a result, we have $|H_i| \geq \chi(T'') = \chi(T')$.
Now recall that for any hyperplane $h$ in $\mathbb{R}^k$, there are at most $k$ points among $r_1,\dots,r_n$ whose distances to $h$ are at most $\varepsilon'$.
This implies that
\begin{equation*}
|H| \geq \sum_{i=1}^{n}\frac{|H_i|}{k} \geq \frac{n\chi(T')}{k}.
\end{equation*}
Therefore, $\chi(T)$ is $\Omega(nN^{k-1})$.
Note that the number of the blue points in $T$ is now $2N+(k+1)n$.
To make it exactly $N$, we only need to choose $N_0 = N/(k+3)$, and use the same method to construct a dataset $T$ containing $n$ red points and $2N_0+(k+1)n$ blue points in general position with $\chi(T) = \Omega(nN_0^{k-1}) = \Omega(nN^{k-1})$.
Then by adding some dummy blue points, we eventually have $T \in \mathcal{T}_{n,N}^d$ with $\chi(T) = \Omega(nN^{k-1})$, which completes the proof.

\subsection{Proof of Theorem \ref{th4}}
Let $(r,b)$ be any closest pair of points where $r \in \mathcal{CH}(T_R)$ and $b \in \mathcal{CH}(T_B)$.
Also, let $h$ be the bisector of the segment $\overline{rb}$.
Then $M_h(T)=\mathit{dist}(r,b)/2$. 
Consider any other separator (of $T$) $h' \neq h$.
We have that

\begin{equation*}
\min\{\mathit{dist}(r,h'),\mathit{dist}(b,h')\}<\mathit{dist}(r,b)/2.
\end{equation*}

Furthermore, since $r \in \mathcal{CH}(T_R)$ and $b \in \mathcal{CH}(T_B)$,
$M_{h'}(T)$ must be less than or equal to $\min\{\mathit{dist}(r,h'),\allowbreak \mathit{dist}(b,h')\}$.
Thus,
\begin{equation*}
M_{h'}(T) \leq \min\{\mathit{dist}(r,h'),\mathit{dist}(b,h')\} < \mathit{dist}(r,b)/2 = M_h(T).
\end{equation*}
This implies that $h$ is the unique maximum-margin separator, 
though the closest pair $(r,b)$ may be not unique.

\subsection{Proof of Theorem \ref{th5}}
Let $h$ be the maximum-margin separator of $T$ and $M = \mathit{Mar}(T)$ be the separation-margin of $T$.
Also, let $(r,b)$ be the closest pair where $r \in \mathcal{CH}(T_R)$ and $b \in \mathcal{CH}(T_B)$.
From the proofs of Theorem \ref{th1} and \ref{th4}, we know that $\mathit{dist}(r,h)=\mathit{dist}(b,h)=M$.
It immediately follows that $r \in \mathcal{CH}(C_R)$ and $b \in \mathcal{CH}(C_B)$.
Since $\mathcal{CH}(C_R) \subseteq \mathcal{CH}(T_R)$ and $\mathcal{CH}(C_B) \subseteq \mathcal{CH}(T_B)$, $(r,b)$ is also the closest pair w.r.t. $\mathcal{CH}(C_R)$ and $\mathcal{CH}(C_B)$.
Thus, both the maximum-margin separator and separation-margin of $C$ are the same with those of $T$.
Furthermore, because all of the points in $C$ have the same distance to $h$, the support set of $C$ is just $C$ itself.

\subsection{Proof of Theorem \ref{th6}} \label{proof:th6}
The number of possible support sets of size smaller than or equal to $d$ is clearly bounded by $O(nN^d)$.
So we only need to bound the number of the ones of sizes are larger than $d$.
We first arbitrarily label all the points in $S$ from 1 to $(n+N)$.
For any $C \subseteq S$ with $|C|>d$, define the \textit{representation} of $C$ as the $(d+1)$ points in $C$ with the smallest labels (we say those $(d+1)$ points represent $C$).
Let $a_1,a_2,\dots,a_{d+1}$ be a tuple of $(d+1)$ points in $S$ where $a_1,\dots,a_k$ are red and $a_{k+1},\dots,a_{d+1}$ are blue.
If $k=0$ or $k=d+1$, there is no possible support set represented by these $(d+1)$ points because the number of the blue/red points in the support set can at most be $d$.
Now consider the case that $1 \leq k \leq d$.
It is easy to see that there exist a unique pair of parallel hyperplanes $(h_r,h_b)$ such that $h_r$ goes through $a_1,\dots,a_k$ and $h_b$ goes through $a_{k+1},\dots,a_{d+1}$, as $S$ is in general position.
If a possible support set $C$ is represented by $a_1,a_2,\dots,a_{d+1}$, then $h_r$ and $h_b$ must be the corresponding support planes.
That means all the red/blue points in $C$ must lie on $h_r$/$h_b$.
Note that there are at most $2d$ points on $h_r$ and $h_b$, which implies that the number of such $C$ is constant.
Since the number of such $(d+1)$-tuples is $O(nN^d)$, $S$ can have at most $O(nN^d)$ possible support sets.
Since the separation-margin is uniquely determined by support set, the number of the possible separation-margins is 
also bounded by $O(nN^d)$.

\subsection{Proof of Theorem \ref{tight}}
First, we define $(d+1)$ probability distributions $D_0,D_1,\dots,D_d$ in $\mathbb{R}^d$, where $D_i$ is a uniform distribution over an $\varepsilon$-ball ($\varepsilon$ is a small enough positive constant) centered at $c_i$.
Set
\begin{equation*}
c_0 = (0,\dots,0),
\end{equation*}
\begin{equation*}
c_i = (\underbrace{0,\dots,0}_{i-1},1,\underbrace{0,\dots,0}_{d-i}),\ \text{for } i = 1,\dots,d.
\end{equation*}
Now we randomly generate a stochastic dataset $S^* = S_R^* \cup S_B^*$ with $n$ red points and $N$ ($N \geq n$) blue points as follows.
The existence probabilities of all the bichromatic points are set to be 0.5.
The location of each red point is drawn from the distribution $D_0$.
For the blue points, we evenly separate them into $d$ groups each of which contains $N/d$ points (for convenience, assume $N$ is a multiple of $d$).
Then the location of each blue point in the $i$-th group is drawn from the distribution $D_i$.
All the locations are drawn independently.
We claim that the randomly generated $S^*$ satisfies
\begin{equation*}
\Pr\left[\ \kappa(S^*) \geq n \left( \frac{N}{d} \right)^d\ \right] > \delta
\end{equation*}
for any $\delta < 1$, whence the existence of $S$ with $\kappa(S) = \Theta(n N^d)$ is shown.
We denote the points in $S_R^*$ by $r_1,\dots,r_n$ and the points in $S_B^*$ by $b_1,\dots,b_N$.
Also, we denote by $B_i$ the subset of $S_B^*$ which contains all the points drawn from $D_i$.
Consider all the $(d+1)$-tuples $(r_j,b_{\pi_1},\dots,b_{\pi_d})$ where
\begin{equation*}
r_j \in S_R^*,\ \ b_{\pi_i} \in B_i.
\end{equation*}
Clearly, we have in total $n(N/d)^d = \Theta(nN^d)$ such tuples.
For each tuple $\tau = (r_j,b_{\pi_1},\dots,b_{\pi_d})$, define $C_\tau = \{r_j,b_{\pi_1},\dots,b_{\pi_d}\}$.
We show that $\xi_{S^*}(C_\tau) > 0$, i.e., $C_\tau$ is a possible support set.
Let $h_b$ be the hyperplane goes through $b_{\pi_1},\dots,b_{\pi_d}$.
Since $\varepsilon$ is small enough, from the spatial locations of $D_0,D_1,\dots,D_d$, we observe that the point on $h_b$ closest to $r_j$ is always inside the simplex $(b_{\pi_1} \dots b_{\pi_d})$, no matter what the exact locations of $r_j,b_{\pi_1},\dots,b_{\pi_d}$ are.
It follows that $C_\tau$ is the support set of the instance of $S^*$ in which the only existent points are those in $C_\tau$.
Thus, $\xi_{S^*}(C_\tau) > 0$.
Define $M_\tau = \mathit{Mar}(C_\tau)$, i.e., the separation-margin of $C_\tau$.
It is easy to see that, for any two distinct tuples $\tau$ and $\tau'$, the probability of $M_\tau = M_{\tau'}$ is infinitesimal, i.e.,
\begin{equation*}
\Pr [\ M_\tau = M_{\tau'}\ ] < c
\end{equation*}
for any small $c>0$.
Therefore, we have
\begin{equation*}
\Pr [\ M_\tau \neq M_{\tau'} \text{ for } \forall \tau \neq \tau'\ ] \geq 1 - \sum_{\tau \neq \tau'} \Pr [\ M_\tau = M_{\tau'}\ ] > \delta
\end{equation*}
for any $\delta < 1$.
Note that if all $M_\tau$ are distinct, $\kappa(S^*)$ is at least $n(N/d)^d$.
So we can conclude
\begin{equation*}
\Pr\left[\ \kappa(S^*) \geq n \left( \frac{N}{d} \right)^d\ \right] \geq \Pr [\ M_\tau \neq M_{\tau'} \text{ for } \forall \tau \neq \tau'\ ].
\end{equation*}
As a result, there exists some stochastic dataset $S$ with $\kappa(S) = \Theta(n N^d)$.

\subsection{Proof of Theorem \ref{th3ball}} \label{th3ball-proof}
By applying the same method used in the proof of Theorem \ref{th3} (see Appendix~\ref{th2-proof}), we can directly show that $T^{U^*}$ is weakly separable.
However, to prove the remaining part, we need to slightly change the method in the proof of Theorem \ref{th3}.
First, we modify the definition of ``degree'' as follows.
Let $X$ be the convex hull of a finite set of balls and $x$ be a point on the boundary of $X$.
Also, let $Y$ be the union of those balls.
We define the degree of $x$ in $X$, denoted by $\deg_Xx$, to be the minimum of the dimensions of all the simplices that contain $x$ and use only the points in $Y$ as their vertices.
We use $C_1$ and $C_2$ to denote the convex hulls of the balls in $T_R^{U^*}$ and $T_B^{U^*}$, respectively.
Since $T^{U^*}$ is not strongly separable, there exists a point $x^* \in C_1 \cap C_2$.
Suppose $\deg_{C_1}x^* = k_1$ and $\deg_{C_2}x^* = k_2$.
Then we can find a $k_1$-dim (resp. $k_2$-dim) simplex $\bar{s}_R$ (resp. $\bar{s}_B$) satisfying \\
\textbf{1)} $\bar{s}_R$ (resp. $\bar{s}_B$) contains $x^*$ in its interior; \\
\textbf{2)} each vertex of $\bar{s}_R$ (resp. $\bar{s}_B$) is contained in at least one ball in $T_R^{U^*}$ (resp. $T_B^{U^*}$). \\
Consider the balls that contain the vertices of $\bar{s}_R$ and $\bar{s}_B$.
We have two cases.
First, all of those balls are 0-radius balls.
Second, at least one of them has the radius larger than 0.
For the first case, the proof of Theorem \ref{th3} is sufficient to show that the weak separator of $T^{U^*}$ is unique and goes through $d$ points (0-radius balls).
In the second case, without loss of generality, we assume there is a vertex of $\bar{s}_R$, $v$, contained in a ball $B \in T_R^{U^*}$ with radius larger than 0.
Since any weak separator of $T^{U^*}$ must go through $v$, $v$ must be on the boundary of $B$.
Thus, $T^{U^*}$ has only one weak separator, which is the tangent hyperplane of $B$ on $v$ (so it is tangent to at least one ball with radius larger than 0).

\subsection{Proof of Theorem \ref{th7}}
Let $P_0$ and $P_1$ be the pre-images of $\{0\}$ and $\{1\}$ under the map $\pi_T^*$ respectively.
Also, let $P_0'$ and $P_1'$ be the counterparts under the map $\pi_C^*$.
Suppose $u^*$ is the clockwise boundary of $P_0$.
Since $C \subseteq T$, we have $P_0' \subseteq P_0$.
On the other hand, as $C$ is the critical set of $T$, it is easy to see that $\mathcal{CH}(C_R^{U^*}) \cap \mathcal{CH}(C_B^{U^*}) \neq \emptyset$, where $U^* = \sigma(u^*)$.
This then implies $u^* \in P_0'$.
Now because $P_0'$ is nonempty, the extreme separator of $C$ is directly defined.
Furthermore, from the fact that $u^* \in P_0' \subseteq P_0$, we know $u^*$ is also the clockwise boundary of $P_0'$ so that $U^*$ is the auxiliary subspace of both $T$ and $C$.
To prove $T$ and $C$ share the same extreme separator, we assume $h$ is the unique weak separator of $T^{U^*}$.
Since $C^{U^*} \subseteq T^{U^*}$, $h$ is also a weak separator of $C^{U^*}$.
More precisely, $h$ is the unique weak separator of $C^{U^*}$, as $C^{U^*}$ only has one weak separator (according to Theorem \ref{th3ball}).
Consequently, the derived separator of $h$ in $\mathbb{R}^d$ is the extreme separator of both $T$ and $C$.

\section{A deterministic algorithm to compute \texorpdfstring{$\mathcal{A}$}{A}} \label{append:compute_matrix_A}
Given a set $S = \{s_1, s_2, \dots, s_n\}$ of $n$ points in general position, where $s_i \in \mathbb{R}^d$,
we propose a deterministic algorithm that computes, in $O(n^{d-1})$ time, a $d \times d$ orthogonal matrix $\mathcal{A} = (\mathbf{a}_1, \mathbf{a}_2, \dots, \mathbf{a}_d)^T$ which linearly transforms $S$ into a new set $S' = \{\mathcal{A}s_1, \mathcal{A}s_2, \dots, \mathcal{A}s_n\}$ satisfying SGPP.
According to the definition of SGPP, what we want is that, for any $k \in \{0,1,\dots, \lfloor (d - 1) / 2 \rfloor\}$, $\Phi_{\{2k+1, 2k+2, \dots, d\}}(S')$ is in general position in $\mathbb{R}^{d - 2k}$.
For $k=0$, $\Phi_{\{2k+1, 2k+2, \dots, d\}}(S') = S'$ is for sure in general position if $\mathcal{A}$ is orthogonal.
For $k \geq 1$, it is easy to see that $\Phi_{\{2k+1, 2k+2, \dots, d\}}(S')$ is in general position iff
\begin{equation}\label{eqn:span_Rd}
\text{dim}(\text{span}\{\mathbf{a}_1, \mathbf{a}_2, \dots, \mathbf{a}_{2k-1}, \mathbf{a}_{2k}, (\mathbf{s}_{i_1} -\mathbf{s}_{i_2}), 
(\mathbf{s}_{i_1} - \mathbf{s}_{i_3}), \dots, 
(\mathbf{s}_{i_1} - \mathbf{s}_{i_{d-2k+1}})\}) = d,
\end{equation}
for any distinct $(d - 2k + 1)$ points $s_{i_1}, \dots, s_{i_{d-2k+1}} \in S$.
Based on this fact, we show how to compute $\mathbf{a}_{2k+1}$ and $\mathbf{a}_{2k+2}$ at a time as $k$ increases from 0.
(This also implies that $\mathbf{a}_1, \dots, \mathbf{a}_{2k}$ are already given when computing $\mathbf{a}_{2k+1}$ and $\mathbf{a}_{2k+2}$.)

For a particular $k$, we first find a candidate $\mathbf{a}_{2k+1}$ satisfying
$$\text{dim}(\text{span}\{\mathbf{a}_1, \mathbf{a}_2, \dots, \mathbf{a}_{2k+1}, 
(\mathbf{s}_{i_1} - \mathbf{s}_{i_2}), 
(\mathbf{s}_{i_1} - \mathbf{s}_{i_3}), \dots, 
(\mathbf{s}_{i_1} - \mathbf{s}_{i_{d-2k-1}})\}) = d - 1,$$
for any distinct $s_{i_1}, \dots, s_{i_{d-2k-1}} \in S$.
In other words, the candidate $\mathbf{a}_{2k+1}$ cannot lie in any $(d - 2)$-dim subspace, $V$, spanned by 
$\{\mathbf{a}_1, \mathbf{a}_2, \dots, \mathbf{a}_{2k},
(\mathbf{s}_{i_1} - \mathbf{s}_{i_2}), 
(\mathbf{s}_{i_1} - \mathbf{s}_{i_3}), \dots, 
(\mathbf{s}_{i_1} - \mathbf{s}_{i_{d-2k-1}})\}$.
Based on this, we propose a simple method that guarantees to find such a candidate $\mathbf{a}_{2k+1}$ as follows.
Initialize an open ball $B$ centered at $c = (1, 1, \dots, 1)$ with radius $r = 0.5$ in $\mathbb{R}^d$.
We enumerate all possible $\mathbf{s}_{i_1}, \dots, \mathbf{s}_{i_{d-2k-1}}$, 
and shrink this ball gradually to a non-empty feasible region for the candidate $\mathbf{a}_{2k+1}$.
Let $\mathbf{s}_{i_1}, \dots, \mathbf{s}_{i_{d-2k-1}}$ be the current enumerated tuple, 
and $B = (c, r)$ be the ball maintained so far.
Let the span $V$ be defined the same as above.
Now, consider two cases: $c \not\in V$ and $c \in V$.
In the first case, we simply reduce $r$ to $r' = \min\{r, \mathit{dist}(c, V)\}$; 
in the second case, we choose an arbitrary point $c' \in B - V$ as the new center of the ball, 
and then set the new radius $r' = \min\{\mathit{dist}(c', V), r - \mathit{dist}(c, c')\}$.
After this shrinking, the new ball is contained in the previous one and does not intersect with the subspace $V$.
In this way, after $\binom{n}{d - 2k - 1} = O(n^{d-2k-1})$ steps of enumeration, 
the final ball indicates a non-empty feasible region for the candidate $\mathbf{a}_{2k+1}$.
We then pick any point in it as our candidate $\mathbf{a}_{2k+1}$.
\begin{figure}[htpb]
	\centering
	\begin{subfigure}{.45\textwidth}
	    \centering
		\includegraphics[height=3cm]{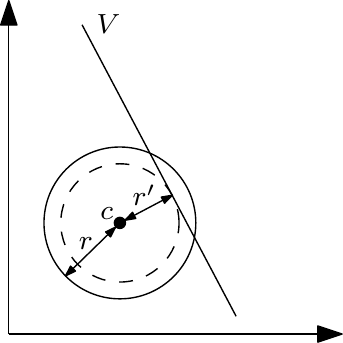}
		\caption{Case 1: $c \not\in V$}
		\label{fig:case1}
	\end{subfigure}
	\begin{subfigure}{.45\textwidth}
	    \centering
		\includegraphics[height=3cm]{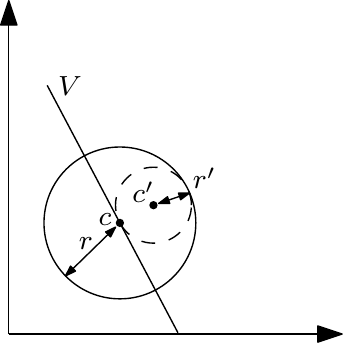}
		\caption{Case 2: $c \in V$}
		\label{fig:case2}
	\end{subfigure}
	\caption{Illustrating two cases in $\mathbb{R}^2$, where the subspace $V$ is a line, and the feasible region is an open disk.
		The unshrunk disk $B$, w.r.t. $V$, is marked in solid line, and the shrunk disk $B'$ is in dash.}
\end{figure}

After the candidate $\mathbf{a}_{2k+1}$ is found, we modify it to guarantee the orthonormality of $\{\mathbf{a}_1,\dots,\mathbf{a}_{2k+1}\}$ without changing its span, and use the vector after modification as the $(2k+1)$-st row vector of $\mathcal{A}$, $\mathbf{a}_{2k+1}$.
Next, the row vector $\mathbf{a}_{2k+2}$ can be computed similarly, i.e.,
first computing a candidate $\mathbf{a}_{2k+2}$ satisfying Equation~\ref{eqn:span_Rd} (note that, in this case, each infeasible region $V'$ is a $(d - 1)$-dim subspace instead) and then modifying it to guarantee the orthonormality.
In this way, determining $\mathbf{a}_{2k+1}$ and $\mathbf{a}_{2k+2}$ takes $O(n^{d-2k-1})$ time,
and the transformation matrix $\mathcal{A}$ can be computed in 
$O(n^{d-1} + n^{d-3} + n^{d-5} + \dots) = O(n^{d-1})$ time.

\section{Computing the separable-probability in \texorpdfstring{$O(nN^{d-1})$}{O(nNd-1)} time}\label{append:improving_prob}
Previously, we showed how to solve the problem by enumerating $d - 1$ points first, followed by a radial-order sort and a sliding windows technique on the remaining points.
This method takes $O(nN^{d - 1} \log N)$ time.
Inspired by \cite{martin2015seperability}, we show how to eliminate the log factor by the well-known techniques of {\it duality} \cite{deBerg_CG_book} and {\it topological sweep} \cite{Edelsbrunner:1986:topological_sweep} as follows.

We first enumerate $d - 2$ points (of which at least one is red and at least one is blue), and these points span a $(d - 3)$-dim subspace $\mathcal{D}$, which corresponds to a 2D dual subspace, $\mathcal{D}^*$.
By duality, each remaining point, $p$, maps to a $(d - 1)$-dim hyperplane, $p^*$, in the dual space, whose intersection with $\mathcal{D}^*$ is a line, $l$.
(Since there is a clear one-to-one correspondence between $p^*$ and $l$, with a slight abuse of notation, we use $p^*$ to represent $l$ below.)
It then follows that there are $n + N - d + 2 = O(N)$ lines in $\mathcal{D}^*$, forming a line arrangement,
and the dual of each intersection point, $f^*$, formed by two lines $p^*_1$ and $p^*_2$ is the span $f$ of some $(d - 1)$-dim facet in the primal space.
We define the \textit{statistic} of $f^*$ as a tuple in form of ($\mathcal{R}^-, \mathcal{R}^+, \mathcal{B}^-, \mathcal{B}^+, \mathcal{T})$, where $\mathcal{R}^-$ and $\mathcal{R}^+$ (resp., $\mathcal{B}^-$ and $\mathcal{B}^+$) denote the non-existence probability of the remaining red (resp., blue) points on either side of $f$, and $\mathcal{T}$ is a set consisting of all the points on $f$.
Given the statistic for every $f^*$, the probability of each $f^*$, i.e., each enumerated facet in the primal plane, can be computed in $O(1)$ time. 
Thus, it suffices to show how to compute the statistics for all $f^*$ efficiently.

\begin{figure}[htpb]
	\centering
	\begin{subfigure}[t]{.47\textwidth}
	    \centering
		\includegraphics{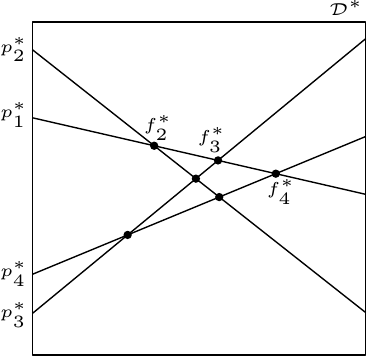}
		\caption{An example of the arrangement in the subspace $\mathcal{D}^*$}
		\label{fig:dual}
	\end{subfigure}
	\begin{subfigure}[t]{.45\textwidth}
	    \centering
		\includegraphics{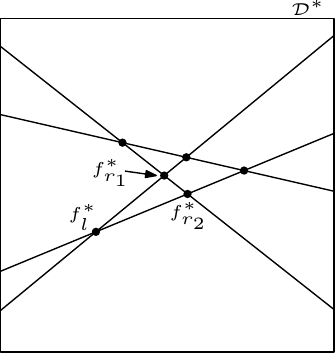}
		\caption{A elementary step in topological sweep}
		\label{fig:dual1}
	\end{subfigure}
	\caption{Illustrating how to use duality and topological sweep to eliminate the log factor.}
\end{figure}

Assume the lines in $\mathcal{D}$ are $p^*_1, \dots, p^*_m$, and the intersection points on $p^*_1$
are $f^*_2, \dots, f^*_m$.
W.l.o.g., assume $f^*_2, \dots, f^*_m$ are sorted from left to right in $\mathcal{D}^*$.
We first compute the statistic for $f_2^*$ by brute-force, which takes $O(N)$ time.
Then, we iterate through $f^*_3, \dots, f^*_m$ from left to right on $p^*_1$.
(See Figure~\ref{fig:dual} for an example.)
By duality, the movement $f^*_{i - 1} \rightarrow f^*_i$ corresponds to 
the hyperplane rotation from $f^*_{i - 1}$ to $f^*_{i}$ w.r.t. the dual of line $p^*_1$,
which is a $(d - 2)$-dim subspace in the primal space.
More importantly, the rotation does not hit any other points except $p_{i-1}$ and $p_i$,
which allows us to quickly compute, in $O(1)$ time, the statistic of $f_i^*$ from that of $f_{i-1}^*$.
This way, the statistics of all the intersections along $p^*_1$ can be computed in $O(N)$ time  without considering the sorting.

In fact, we cannot afford to sort the intersections on each line since that will take $O(N^2 \log N)$ time.
Instead, we compute the entire line arrangement using $O(N^2)$ time and space,
then we can visit the intersections on each line in the correct order (though not necessarily consecutively).
To further reduce the space from $O(N^2)$ to $O(N)$, one can perform a {\it topological sweep} on the arrangement.
The topological sweep maintains a cut of size $O(N)$, and sweep it from left to right over the entire line arrangement using $O(N^2)$ so-called elementary steps, each taking $O(1)$ amortized time. (See Figure~\ref{fig:dual1} for details.)
Based on this, we find the leftmost intersection point, $f^*_l$, in $\mathcal{D}^*$,
and compute statistic of it by brute-force. 
This step takes $O(N^2)$ time.
Afterwards, when an elementary step is triggered, 
the statistic for the current intersection point, $p^*$, can be reported,
and we can compute, in $O(1)$ time, the statistics for two more intersections points 
(e.g., $f^*_{r_1}$ and $f^*_{r_2}$ in Figure~\ref{fig:dual1})
for future reporting.
Thus, as we advance from the leftmost cut to the rightmost one,
the statistics of all the intersection points are reported on the fly.
Therefore, the runtime of our algorithm is improved to $O(nN^{d-3} \cdot N^2) = O(nN^{d-1})$, using linear space.
{\bf Remark.} Note that, in $\mathbb{R}^2$ only, the above method actually runs in $O(N^2)$ instead of $O(nN)$. However, the runtime of our previous method based on radial-order sort still remains $O(nN \log N)$.

\section{Improving the time for computing the expected separation-margin} \label{append:improving_margin}
It is easy to improve the time for computing the ESM to $O(nN^d \log N)$ by slightly modifying the sort method we used for improving our SP computing algorithm.
When enumerating $(d+1)$ points, we first determine $d$ points (of which at least one is red and one is blue).
Let us denote by $r_1,\dots,r_k$ the red ones and $b_1,\dots,b_{d-k}$ the blue ones.
We can uniquely determine two parallel $(d-2)$-dim linear subspaces $X_r$ and $X_b$ such that $r_1,\dots,r_k \in X_r$ and $b_1,\dots,b_{d-k} \in X_b$.
We sort all the remaining red points around $X_r$ and the blue ones around $X_b$.
Then we consider the last point in that sorted order (say red first and then blue) and meanwhile maintain two sliding windows (for red and blue points respectively).
In this way, we are able to use amortized constant time for considering each tuple of $(d+1)$ points, i.e., computing the probabilities of all the possible support sets represented by the $(d+1)$ points and adding the portions contributed by these possible support sets to the ESM.
Thus, the computation of the ESM can be done in $O(nN^d \log N)$ time.

To further improve the runtime to $O(nN^d)$ requires more effort.
We can still apply the duality and topological sweep techniques but the approach is somewhat different from that in the SP problem.
For convenience, we define the \textit{red} (resp., \textit{blue}) \textit{statistics} of a hyperplane $h$ to be the set of the red (resp., blue) points on $h$ and the product of the non-existence probabilities of all the red (resp., blue) points on each side of $h$.
As we see, in the process of computing the SP, the object enumerated is one hyperplane spanned by $d$ points and what we want to compute is the red and blue statistics of the hyperplane.
In this situation, the idea of duality and topological sweep can be directly used to improve the efficiency of each computation.
However, when computing the ESM, the situation is different.
At each step, we have three parallel and equidistant hyperplanes $(h_r,h,h_b)$ determined by $(d+1)$ points, and what we want to compute is the red statistics of $h_r$ and the blue statistics of $h_b$.
Thus, in order to apply the duality and topological sweep techniques, our idea is to transform the problem from the latter form to the former one.
We consider two different cases: $d \geq 3$ and $d = 2$.

Suppose $d \geq 3$.
In this case, when enumerating $(d+1)$ points, we first determine two of them, of which one is red and one is blue (say $r$ and $b$).
Let $c$ be the midpoint of the segment $\overline{rb}$.
Then for each $r_i \in S_R$, we construct a new red point $r'_i = r_i + \overrightarrow{rc}$ with existence probability the same as that of $r_i$.
And for each $b_i \in S_B$, we construct a new blue point $b'_i = b_i + \overrightarrow{bc}$ with existence probability the same as that of $b_i$.
Denote by $S'$ the new stochastic dataset of those constructed points.
Now consider any tuple of $(d+1)$ points in $S$ including $r$ and $b$.
Let $(h_r,h,h_b)$ be the three hyperplanes determined by these $(d+1)$ points.
In order to complete the computation for this tuple, what we need to know is the red statistics of $h_r$ and the blue statistics of $h_b$.
It is easy to see that: \\
$\bullet$ A red (resp., blue) point in $S$ is on $h_r$ (resp., $h_b$) iff its corresponding point in $S'$ is on $h$. (So each of the $(d+1)$ points corresponds to a point on $h$.)
\\
$\bullet$ The red (resp., blue) points in $S$ on each side of $h_r$ (resp., $h_b$) just correspond to the red (resp., blue) points in $S'$ on each side of $h$. \\
Based on the above observations, the red statistics of $h_r$ and the blue statistics of $h_b$ w.r.t. $S$ just correpond to the red and blue statistics of $h$ w.r.t. $S'$.
In other words, to consider all the possible support sets represented by this tuple, it suffices to know the red and blue statistics of $h$ w.r.t. $S'$.
Now the problem we face is similar to that in the SP problem.
We want to compute, for each hyperplane $h$ spanned by the point $c$ and other $(d-1)$ points in $S'$, the red and blue statistics of $h$.
By applying the idea of duality and topological sweep, this can be done in $O(N^{d-1})$ time.
This is the runtime for a fixed $(r,b)$.
To compute the ESM, we need to enumerate all $O(nN)$ such pairs so that the overall time is $O(nN^d)$.

In the case of $d=2$, however, the above method does not work.
Since we enumerate three points when $d=2$, if we first determine two of them ($r$ and $b$), we are not able to create the line arrangement in the dual space and use topological sweep to complete the computation for $(r,b)$ in $O(N)$ time.
So we need to deal with the 2-dim problem separately.
Without loss of generality, we only consider the case where the three points enumerated are one red point and two blue points (the two-red one-blue case is symmetric).
Let $n_r$ be the number of the red points and $n_b$ the number of the blue ones.
When enumerating three points, we first determine the red point $r$ and sort all the other red points around $r$.
Then for all the blue points, we construct their dual lines to form a line arrangement.
Each vertex (i.e., intersection point) of the arrangement corresponds to a pair of blue points $(b_i,b_j)$.
We want to apply topological sweep on the arrangement and consider each 3-tuple $(r,b_i,b_j)$ at the time we visit the vertex $(b_i,b_j)$.
Let $(r,b_i,b_j)$ be any such tuple and $(h_r,h,h_b)$ be the three hyperplanes determined by this tuple.
In order to complete the computation for this tuple, we need to know the red statistics of $h_r$ and the blue statistics of $h_b$.
We note that the hyperplane $h_b$ is actually determined by $b_i$ and $b_j$ only (independent of $r$).
Thus, the blue statistics of $h_b$ can be directly computed in the process of topological sweep.
The crucial part is to compute the red statistics of $h_r$.
What we do is to maintain $n_b$ sliding windows $w_1,\dots,w_{n_b}$ on the sorted list of the red points, where $w_i$ corresponds to the blue point $b_i$.
During the topological sweep, the sliding window $w_i$ dynamically indicates the red points on one side of the hyperplane $h_r$ determined by the tuple $(r,b_i,b^*)$, where $(b_i,b^*)$ is the most recently visited vertex on the dual line of $b_i$.
At each time a new vertex $(b_i,b_j)$ is visited, we update $w_i$ and $w_j$, and meanwhile compute the red statistics of the hyperplane $h_r$ determined by the tuple $(r,b_i,b_j)$.
It is easy to see that both updating the sliding windows and computing the statistics can be done in amortized constant time.
Therefore, for each red point $r$, the computations take $O(n_b^2)$ time.
The total time for considering all the red points is then $O(n_r n_b^2)$, which is bounded by $O(nN^2)$.
Symmetrically, the work for enumerating two red points and one blue point can also be done in $O(nN^2)$ time.

As a result, for any $d \geq 2$, the ESM of a stochastic bichromatic dataset $S$ in $\mathbb{R}^d$ can be computed in $O(nN^d)$ time.

\section{Extension to multipoint model}\label{sec:multipoint}
All our algorithms in the paper can be generalized in a straightforward manner from the unipoint model to the multipoint model with the same bounds.
Let $S = S_R \cup S_B$ be set of stochastic bichromatic points under the multipoint model,
i.e., $S = \{A_1,A_2,\dots,A_{m}\}$, where each 
$A_i = \{(a_1^{(i)},p_1^{(i)}), (a_2^{(i)},p_2^{(i)}), \dots, (a_{n_i}^{(i)},p_{n_i}^{(i)})\}$ represents an uncertain point, where $a_j^{(i)} \in \mathbb{R}^d$ is its $j$-th possible location and $p_j^{(i)} \in (0,1]$ is its corresponding probability of existing at $a_j^{(i)}$.
With a slight abuse of the notation,
let $|S_R|$ (resp., $|S_B|$) be the total number of locations of all red (resp., blue) multipoints, and define $n = \min\{|S_R|, |S_B|\}$ and $N = \max\{|S_R|, |S_B|\}$.
Then the total size of S is $\sum_{i=1}^{m}{n_i} = n + N$.
Clearly, $S$ can be regarded as a unipoint-model stochastic dataset
$\rule{0mm}{5mm} S' = \{(a_j^{(i)},p_j^{(i)}): i \in \{1,\dots,m\}, j \in \{1,\dots,n_i\}\},$
where the existences of $a_1^{(i)},\dots,a_{n_i}^{(i)}$ are dependent (i.e., at most one of them can exist) for $i \in \{1,\dots,m\}$.
Thus, while applying our algorithms under the multipoint model, the only issue is that we need to deal with such dependences.
Note that in our algorithms all the problems are finally transformed into one form: 
computing the probability that some points definitely exist and some other points definitely do not.
Let $\mathcal{X}$ (resp., $\bar{\mathcal{X}}$) be the set containing all the points that are definitely present (resp., absent).
For any uncertain point $A_i$, consider the following three cases.\\
\textbf{1)} If $|A_i \cap \mathcal{X}| \ge 2$, the probability contributed by $A_i$ is 0 since an uncertain point cannot exist at two different places simultaneously.\\
\textbf{2)} If $|A_i \cap \mathcal{X}| = 1$, the probability contributed by $A_i$ is equal to the probability of the only element in $A_i \cap \mathcal{X}$.\\
\textbf{3)} If $|A_i \cap \mathcal{X}| = 0$, the probability contributed by $A_i$ is simply $1 - \sum p_i$, for all
$(a_i, p_i) \in A_i \cap \bar{\mathcal{X}}$.\\
Finally, the probability for the scenario $(\mathcal{X}, \bar{\mathcal{X}})$ is equal to the product of the probabilities contributed by all $A_i$.

In this way, by only slightly modifying the previous way of computation, the dependences among the existences of the points can be easily handled without any increase in the running time.

\end{document}